\documentclass[11pt,a4paper]{amsart}
\usepackage[T1]{fontenc}
\usepackage[latin1]{inputenc}
\usepackage[extra]{tipa}
\usepackage{fancyhdr}
\usepackage{amsmath, amssymb, enumerate, xspace, graphicx, url}
\usepackage[all]{xy}
\SelectTips{cm}{}
\makeindex

\makeatletter
\def\blfootnote{\xdef\@thefnmark{}\@footnotetext}

\numberwithin{equation}{section}

2

\numberwithin{subsection}{section}

\newtheorem*{namedtheorem}{\theoremname}
\newcommand{\theoremname}{testing}

\newtheorem{theorem}{Theorem}[section]
\newtheorem{prop}[theorem]{Proposition}
\newtheorem{proposition-definition}[theorem]
{Proposition-Definition}

\theoremstyle{definition}
\newtheorem{definition}[theorem]{Definition}

\newtheorem{example}[theorem]{Example}

\newtheorem{remark}[theorem]{Remark}

\newtheorem{namedtheoremr}[theorem]{\theoremnamer}
\newcommand{\theoremnamer}{testing}

\theoremstyle{remark}

\newcounter{steps}

\newcommand\cD{\mathcal{D}}

\newcommand\cF{\mathcal{F}}

\newcommand\cM{\mathcal{M}}

\newcommand\cP{\mathcal{P}}

\newcommand\cS{\mathcal{S}}

\newcommand\cX{\mathcal{X}}

\newcommand\bCC{\mathbb{C}}

\newcommand\HH{\mathbb{H}}

\newcommand\KK{\mathbb{K}}

\newcommand\PP{\mathbb{P}}
\newcommand\QQ{\mathbb{Q}}
\newcommand\bRR{\mathbb{R}}

\newcommand\VV{\mathbb{V}}

\newcommand\bZZ{\mathbb{Z}}

\newcommand\rR{\mathrm{R}}

\newcommand\rmd{\mathrm{d}}
\newcommand\rme{\mathrm{e}}

\newcommand\La{\Lambda}
\newcommand\om{\omega}
\newcommand{\ov}[1]{\overline{#1}}

\newcommand{\bw}{\bigwedge}
\newcommand\beq{\begin{equation}}
\newcommand\eeq{\end{equation}}
\newcommand\beqn{\begin{equation*}}
\newcommand\eeqn{\end{equation*}}
\newcommand{\tild}[1]{\widetilde{#1}}
\newlength{\bracewidth}

\newcommand\lmt{\longmapsto}
\newcommand\larr{\longrightarrow}
\newcommand\ra{\rightarrow}
\newcommand{\wek}{\wedge_\KK}

\title{The $E^3/\bZZ_3$ orbifold, mirror symmetry, and Hodge structures of Calabi-Yau type}

\author{Sergio Luigi Cacciatori}
\address{Dipartimento di Scienza e Alta Tecnologia, Universit\`a dell'Insubria, Via Valleggio 11, I-22100 Como, Italia, 
and INFN, Sezione di Milano, Via Celoria 16, I-20133 Milano, Italia}
\email{sergio.cacciatori@uninsubria.it}

\author{Sara Angela Filippini}
\address{Dipartimento di Scienza e Alta Tecnologia, Universit\`a dell'Insubria, via Valleggio 11, I-22100 Como, Italia}
\email{saraangela.filippini@uninsubria.it}

\begin{document}

\begin{abstract}
Starting from the K\"ahler moduli space of the rigid orbifold $Z=E^3/\mathbb{Z}_3$ one would expect for the cohomology of the generalized mirror to be
a Hodge structure of Calabi-Yau type $(1,9,9,1)$. We show that such a structure arises in a natural way from rational Hodge structures on
$\Lambda^3 \mathbb{K}^6$, $\mathbb{K}=\mathbb{Q}[\omega]$, where $\omega$ is a primitive third root of unity. We do not try to identify an
underlying geometry, but we show how special geometry arises in our abstract construction. We also show how such Hodge structure
can be recovered as a polarized substructure of a bigger Hodge structure given by the third cohomology group of a six-dimensional Abelian
variety of Weil type. 
\end{abstract}

\maketitle

\section{Introduction}

\noindent
The $Z:=E^3/\bZZ_3$ orbifold is one of the prototypes of Calabi-Yau (CY) threefolds which appeared in string theory, already in the first studies of
supersymmetric compactifications (\cite{CHSW},\cite{GRV}, \cite{Sh}, \cite{FLT}). It is also the first example of rigid CY manifold, thus resisting to the strictly geometrical
versions of the mirror symmetry conjecture. Generalized mirror constructions have been proposed in order to include rigid manifolds in mirror
conjectures (\cite{CDP}, \cite{BB}, \cite{Se}, \cite{KLS}, \cite{Sch}), and in general involve higher dimensional mirror varieties that cannot have
a geometrical interpretation in compactification theory. \\
The Teichm\"uller covering of the K\"ahler moduli space for the orbifold $Z$ can be identified \cite{FFS} with the symmetric manifold (\cite{Hel},X.2.1)
$$
\cM_{3,3}= 
SU(3,3)/S(U(3) \otimes U(3)) \, .
$$
This is one of the four exceptional cases in the classification of homogeneous symmetric special K\"ahler manifolds \cite{Cremmer:1984hc}.
The special K\"ahler geometry of this manifold has been studied in \cite{FFS}. The same special geometry is expected to be associated
to the middle cohomology of any hypothetical geometrical mirror. The aim of this paper is to present an {\em abstract} version
of the problem of constructing a geometrical mirror manifold. With {\em abstract} we mean that we will not worry here about providing a
construction of a manifold having the right cohomology group, but will only look at the properties of the cohomology group itself.
For the cohomology of the mirror of $Z$ one expects a Hodge structure of CY-type $(1,9,9,1)$. We will show that such a structure
can be obtained in a natural way by starting from rational Hodge structures on $\Lambda^3 \KK^6$, where $\KK= \mathbb{Q}[\omega]$ and $\omega$ is a
primitive third root of unity.
In particular, after suitably parameterizing it, we will show that $\cM_{3,3}$ can be identified with the moduli space of
Hodge structures of weight $3$ on $\Lambda^3 (\mathbb{Q}[\omega])^6$, which are indeed Hodge structures of CY-type $(1,9,9,1)$. Next, we will
show how the special geometry arises in the context of this abstract construction and will apply it to the concrete computation
of the holomorphic prepotential. We will never need to use any property of a hypothetical family of varieties having the given abstract
Hodge structure as cohomology. Indeed, we will not try here to further investigate the geometry underlying our constructions, thus remaining
at an algebraic level. However, we will show that the Hodge structure of CY-type $(1,9,9,1)$ can be recovered as a polarized substructure
of a bigger Hodge structure given by the third cohomology group of a six-dimensional Abelian variety of Weil type. This may be considered
as a first step for investigating the geometrical counterpart of these Hodge structures. A more complete analysis is left to a future work.
\\

\noindent
\section{The moduli space $SU(3,3)/ S(U(3) \times U(3))$}\label{sec:moduli}
\noindent
First we decribe the connection between $SU(3,3)$, the total space appearing in the symmetric manifold $\cM_{3,3}$, and the Narain group $SO(6,6)$. In \cite{FFS} another identification between these two groups was given. Then we provide two descriptions of $\cM_{3,3}$, which will appear in the construction of the variation of the Hodge structure associated to this manifold.\\

\noindent
The total space $SU(3,3)$ of the symmetric manifold $\cM_{3,3}$ is strictly related to $SO(6,6)$, the Narain group associated to the Narain lattice describing the string states on a $T^6\simeq E^3$ torus. \\
Indeed, let $J$ be a complex structure on $\bRR^{2n}$. Consider a symmetric positive definite $\bRR$-bilinear form $g$ on $\bRR^{2n}$ which is $J$-invariant,
i.e.\ $g(Jv,Jw) = g(v,w)$, $\forall v, w \in \bRR^{2n}$. We can identify $ (\bRR^{2n}, J) \simeq \bCC^n$.
On $\bCC^n$ we can define a Hermitian form $H$ as
\beq H(v,w) := H_g(v,w)= g(v,w) + i g(v,J w) \, . \label{Hdef} \eeq
Indeed, it is obviously $\bRR$-bilinear and one can easily check that
$H(Jv,w) = i H(v,w)$ and $H(w,v) = \ov{H(v,w)}$. Moreover $H$ is positive definite:
$$H(v,v) = g(v,v) > 0 \quad \mbox{for } v \neq 0 \, .$$
The converse is also true, since to every Hermitian form $H$ one can associate a symmetric positive definite $\bRR$-bilinear form $g$ defined by the real part
of $H$: $g(v,w) = \mathrm{Re} H(v,w)$. \\
Since $H$ is positive definite, we can identify
\beq
SU(n) \simeq SU(H) =\{A \in M(n, \bCC) \, : \, H(Av, Aw) = H(v,w)\} \, .
\eeq
Similarly, $SO(2n)$ is the group of matrices which preserves $g$
\beq
SO(2n)  \simeq SO(g) =\{A \in M(2n, \bRR) \, : \, g(Av, Aw) = g(v,w)\} \,
\eeq
The relation among $H$, $J$ and $g$ implies
\beq
\{A \in SO(2n) \, : \, AJ=JA\} \simeq SU(n) \, .
\eeq
Now, we can weaken the positivity condition and prove that, if the $J$-invariant symmetric form $g$ on $\bRR^{2n}$ is of type $(2p,2q)$, then the corresponding
Hermitian form $H$ on $\bCC^n$ is of type $(p,q)$, where $p+q=n$. We will work with the case of interest $p=q=3$, but the general case follows easily.
To show that if $g$ is of type $(6,6)$ and $J$-invariant (i.e.\ $J \in SO (g)$), then $H$ is of type $(3,3)$, we have to find a decomposition of
$(\bRR^{12}, J) \simeq \bCC^6$ such that
\beq
\begin{array}{ccc}
(\bRR^{12}, J) & \simeq & \bCC^6 \\
\Vert & & \Vert \\
\begin{array}{ccc}
\bRR^6 & \oplus_{\perp_g} & \bRR^6 \\
\tiny{g>0} && \tiny{g<0} \end{array}
& &
\begin{array}{ccc}
\bCC^3 & \oplus_{\perp_H} & \bCC^3  \\
\tiny{H_g>0} && \tiny{H_g<0} \, . \end{array}
\end{array}
\eeq
Take $e_1 \in \bRR^{12}$ such that $g(e_1,e_1)>0$, and set $e_2:= J e_1$ $\ra J e_2 = J^2 e_1=-e_1$. Hence $J$ is represented by the matrix
$\begin{bmatrix} 0 & 1 \\
-1 & 0 \end{bmatrix}$ on the subspace $\langle e_1, e_2 \rangle$.
Then $g(e_2, e_2) = g(J e_1, Je_1)=  g(e_1,e_1)>0$. Since $g(e_1, Je_1)= g(J e_1, J^2 e_1) = - g (J e_1, e_1) = - g(e_1, J e_1)$, we get $g(e_1, e_2)= 0$.
As $g$ is of type $(6,6)$, its restriction to  $\langle e_1, e_2 \rangle^\perp$ is of type $(4,6)$. Hence we can choose
$e_3 \in \langle e_1, e_2 \rangle^\perp$ such that $g(e_3,e_3)>0$. We can set $e_4:= J e_3$ and repeat the same procedure, until one gets six vectors
$$\begin{array}{cccccc}
e_1, & e_2 , & e_3, & e_4, & e_5, & e_6 \\
 & \shortparallel &  & \shortparallel & & \shortparallel \\
 & J e_1 &  & Je_3 & & Je_5
 \end{array}$$
which (after proper normalization) satisfy
$$g(e_i,e_j) = \delta_{i j}\, , \quad \forall i,j=1, \dots, 6.$$
Hence these vectors span an $\bRR^6$ on which $g$ is positive definite, whereas on $(\bRR^6)^\perp = \langle e_1, \dots, e_6 \rangle^\perp$, $g$ will be
negative definite. Hence we can find $e_7 \in  \langle e_1, \dots, e_6 \rangle^\perp$ such that $g(e_7,e_7)<0$. Set $e_8:= J e_7$.
Then $g(e_8,e_8)<0$ and $g(e_7,e_8)=0$. We proceed analogously to build $e_7, \dots, e_{12}$, which extend $e_1, \dots, e_6$ to a basis of $\bRR^{12}$,
on which $J$ and $g$ are represented by
\beq
J = \begin{bmatrix}
\begin{bmatrix} 0 & 1 \\
-1 & 0 \end{bmatrix} & \ldots & 0 \\
\vdots  & \ddots & \vdots \\
0 & \ldots & \begin{bmatrix} 0 & 1 \\
-1 & 0  \end{bmatrix}
\end{bmatrix}
\, , \quad
g =\begin{bmatrix}
\begin{bmatrix} 1 & &   \\
& \ddots & \\
&&1
\end{bmatrix} & 0 \\
0 & \begin{bmatrix}
-1 && \\
& \ddots & \\
& & -1
\end{bmatrix}
\end{bmatrix}
\eeq
A $\bCC$-basis of $(\bRR^{12}, J)  \simeq  \bCC^6$ is given by $e_1, e_3, e_5, e_7, e_9, e_{11}$. On this basis,
{\renewcommand{\arraystretch}{1.7}
$$\begin{array}{rcl}
H(e_i,e_j) &=& g(e_i, e_j) + i g(e_i, J e_j) \\
&=& g(e_i, e_j) \\
&=& \left\{ \begin{array}{l}
\delta_{i j} \mbox{ for } i, j = 1,3,5 \, , \\
- \delta_{i j} \mbox{ for } i, j = 7,9,11\, ,
\end{array}
\right.
\end{array}$$}\\
since $g(e_i, J e_j)=g(e_i, e_{i+1})=0$ by construction. \\
Hence $H$ is positive definite on the subspace $\langle e_1, e_3, e_5 \rangle \simeq \bCC^3$ and negative definite on
$\langle e_7, e_9, e_{11} \rangle \simeq \bCC^3$, so that is of type $(3,3)$.\\

\noindent
We now give a geometrical description of $SU(3,3)/ S(U(3) \times U(3))$.
Let $H$ be a hermitian form of signature $(3,3)$ on $V_\bCC \cong \bCC^6$. We can obviously choose a basis $\{e_1, \dots, e_6\}$ of $V_\bCC$
such that
$$H(z,w) = z_1\ov{w}_1 +  z_2\ov{w}_2 + z_3\ov{w}_3 - (z_4\ov{w}_4 + z_5\ov{w}_5 + z_6\ov{w}_6) \, .$$
Let
\beq
\begin{array}{rcl}
SU(3,3) &=& SU(H) \\
&=&  \{A \in GL_6(\bCC) : H(Az,Aw)=H(z,w) \mbox{ for } z,w \in \bCC^{6}, \det A = 1\}
\end{array} \eeq
be the group of special isometries for $H$ and set
$$\cS_G = \{W \subset \bCC^6 : W \cong \bCC^3, H_{|_W} > 0 \} \, .$$ Then

\begin{prop}
\beq \cS_G \simeq  SU(3,3)/ S(U(3) \times U(3)) \, . \eeq
\end{prop}

\noindent
\begin{proof}
To prove our assertion we will first show that $SU(3,3)$ acts transitively on $\cS_G$ and then we will show that the
stabilizer $\mathrm{Stab}_{SU(3,3)}(W)$ of any $W \in \cS_G$ is isomorphic to $S(U(3) \times U(3))$.\\
Note that, if $A \in SU(3,3)$, $W \in \cS_G$, then $A W \in \cS_G$. Given $W,W' \in \cS$, there exists an $A \in SU(3,3)$
such that $AW = W'$. Indeed, choose orthonormal bases (w.r.t.\ $H$) $\{f_1,f_2,f_3\}$ of $W$ and $\{f'_1,f'_2,f'_3\}$ of $W'$. These can be extended
to orthonormal bases $\{f_i\}_{i=1}^6$ and $\{f'_i\}_{i=1}^6$ of $(W')$ of $V_\bCC \cong \bCC^6$, such that $H$ is given by
$H(z,w) = z_1\ov{w}_1 +  z_2\ov{w}_2 + z_3\ov{w}_3 - (z_4\ov{w}_4 + z_5\ov{w}_5 + z_6\ov{w}_6)$ on both bases.
Hence the map $A$ sending $\{f_i\}$ to $\{f'_i\}$, $i=1, \dots, 6$, is in $U(3,3)$. After multiplication by a suitable constant,
$A$ will be in $SU(3,3)$ and $AW = W'$, proving transitivity.\\
In order to compute the stabilizer of $W_0 = \left\langle e_1, e_2, e_3\right\rangle \in \cS$, note that
$A W_0 = W_0$ implies that $A W_0^\perp =  W_0^\perp$. This yields the following decomposition $$\bCC^6 =  W_0 \oplus (W_0)^\perp$$ and
$A = \left( \begin{array}{ccc}
A' & 0\\
0 & A''\\
\end{array} \right)
$, where $A' ,A'' : \bCC^3 \ra \bCC^3$. Obviously $A',A''$ preserve $H$, so that $A' \in U(W_0) \cong U(3)$ and $A'' \in U(W_0)^\perp \cong U(3)$,
hence $A = (A',A'') = U(3)\times U(3)$. As $SU(3,3) \cap (U(3) \times U(3))=S(U(3) \times U(3))$, we get the assert.
\end{proof}

\noindent 
A second characterization of the symmetric space $SU(3,3)/ S(U(3) \times U(3))$ can be obtained as follows.
Let $V=W \oplus W^\perp$, as before, with $W \in \cS_G$. Assume $\tilde W\in \cS_G$ is generated by $v_1, v_2, v_3$, with $v_i \in V$. We can write
\renewcommand{\arraystretch}{1.6}{$$
\begin{array}{rcl}
\tilde W=\left\langle v_1, v_2, v_3 \right\rangle =
\left\langle\left( \begin{array}{c}
v_1' \\
v_1''
\end{array} \right) \, ,
\left( \begin{array}{c}
 v_2'  \\
 v_2''
\end{array} \right) \, ,
\left( \begin{array}{c}
v_3' \\
v_3''
\end{array} \right)\right\rangle = \left\langle\left( \begin{array}{ccc}
v_1' & v_2' & v_3' \\
v_1'' &  v_2'' & v_3''
\end{array} \right)\right\rangle,
\end{array}$$}
\noindent with $v'_i \in W$ and $v''_i\in W^{\perp}$.
Then, necessarily $\det(v_1' , v_2',  v_3' ) \neq 0$.
Set
$$B:=  \left(
v_1' \, v_2' \, v_3'
\right) \in U(3) \, ,$$
so that
$$\left( \begin{array}{ccc}
v_1' & v_2' & v_3' \\
v_1'' &  v_2'' & v_3''
\end{array} \right) = \left( \begin{array}{c}
B \\
B'
\end{array} \right) \, .$$
Hence
\renewcommand{\arraystretch}{1.6}{$$
\tilde W=  \left\langle \left( \begin{array}{c}
B \\
B'
\end{array} \right)  \right\rangle = \left\langle
\left( \begin{array}{c}
I \\
Z_{\tilde W}
\end{array} \right) \cdot B  \right\rangle =
\left\langle\left( \begin{array}{ccc}
1 & 0 & 0\\
0 & 1 & 0\\
0 & 0 & 1\\
z_{11} & z_{12} & z_{13} \\
z_{21} & z_{22} & z_{23} \\
z_{31} & z_{32} & z_{33}
\end{array} \right) \right\rangle \, , $$}
\\
\noindent
for some complex $3 \times 3$ matrix $Z=Z_{\tilde W}=(z_{i j})$.

\noindent
Since $H_{|_{\tilde W \times \tilde W}} > 0$,
\renewcommand{\arraystretch}{1.6}{$$
\begin{array}{rl}
0 < & (I \quad {}^t \ov{Z}) \left( \begin{array}{ccc}
I & 0\\
0 & -I\\
\end{array} \right)
\left( \begin{array}{c}
I \\
Z
\end{array} \right)  =  I - {}^t \ov{Z} Z \, .
\end{array}
$$
\noindent
Conversely, given $Z \in M(3, \bCC)$ such that $I - {}^t \ov{Z} Z  > 0$, we recover $W_Z := \left\langle \left( \begin{array}{c}
I \\
Z
\end{array} \right) \right\rangle$, the three-dimensional complex subspace identified by $Z$. \\
\noindent
Thus, if we define 
$$ \cS_M = \{Z \in M(3, \bCC) \, : \,  I - {}^t \ov{Z} Z  > 0\} \, , $$
we get a bijection
\beq \begin{array}{rcl}
\cS_G & \longleftrightarrow & \cS_M \\
W=W_Z  &\longleftrightarrow & Z=Z_W \, .
\end{array} 
\eeq
The action of $M =  \left( \begin{array}{ccc}
A & B \\
C & D \\
\end{array} \right)  \in SU(3,3)$ on $\cS_M$ is given by:

\beq
\left\langle \left( \begin{array}{ccc}
A & B \\
C & D \\
\end{array} \right)
\left( \begin{array}{c}
I \\
Z
\end{array} \right) \right\rangle  = \left\langle\left( \begin{array}{c}
A + B Z \\
C + D Z
\end{array} \right)\right\rangle=\left\langle \left( \begin{array}{c}
I \\
(C + D Z)(A + B Z )^{-1}
\end{array} \right) \right\rangle,
\eeq
so that
\beq \begin{array}{rcl}
SU (3,3) \times \cS_M & \larr & \cS_M \\
(M,Z) & \lmt & M \cdot Z = (C + D Z)(A + B Z )^{-1}.
\end{array}
\eeq
Therefore we get the following isomorphisms:
\beq SU(3,3)/ S(U(3) \times U(3)) \simeq \cS_G  \simeq \cS_M  \, . \eeq
\\
The identification of $SU(3,3)/ S(U(3) \times U(3))$ with $\cS_M$ is a key-step in the construction of the variation of the Hodge structure associated
to $SU(3,3)/ S(U(3) \times U(3))$.\\

\noindent
\section{Rational Hodge structures of CY type}\label{sec:rational}
\noindent
The most natural way to obtain the Hodge structure of CY type $(1,9,9,1)$ is to study an abelian variety of Weil type (\cite{BL}, 17.6), since in this framework a Hodge
structure is already at hand. Nevertheless we postpone this computation to the appendix.\\
Instead, here we prefer to obtain the Hodge structure algebraically. In this case, the question whether the constructed Hodge structure corresponds to the
third cohomology of a (possibly CY) threefold remains open. The advantage is that we can get a quite general result. The detailed analysis of the geometrical
viewpoint will be presented in a future paper aimed at realizing a geometrical mirror of the rigid orbifold under consideration.\\

\noindent
\subsection{The $\KK$-vector space $W$}
Let $\KK=\QQ[\om]$, where $\om$ is a primitive third root of unity. Then $\KK^6 = (\QQ[\om])^6$ is a $6$-dimensional $\KK$-vector space. On
$\KK^6 \otimes_\QQ \bRR = \bCC^6$ we can define a complex structure $J$, i.e.\ an $\bRR$-linear map $J: \bCC^6 \ra \bCC^6$ such that $J^2 = - I$. \\
The complex structure $J$ gives a decomposition of $\KK^6 \otimes_\QQ \bRR = \bCC^6$:
\beq
J: \bCC^6 \larr \bCC^6 \, , \quad \bCC^6 = V_J^{1,0} \oplus V_J^{0,1} \, ,
\eeq
such that
\beq
\label{eigensp}
J v = i v \, , \quad J v' = -i v' \, , \quad \mbox{for every } v \in V_J^{1,0}, v' \in V_J^{0,1} \, .
\eeq
$J$ extends to a $\bCC$-linear map since it commutes with multiplication by $i$. This defines a representation of the abelian group $\bCC^*$ on $\KK^6$ by
\beq
\begin{array}{rrcl}
h: & \bCC^* & \larr & GL(\KK^6 \otimes_\QQ \bRR) \\
& a+b i & \lmt & h(a+b i ):= a \mathrm{Id} + b J \, .
\end{array}
\eeq
The action of $\KK$ on $\KK^6$ is given by multiplication by $\om$, hence the action of $\KK$ and $J$ commute.\\
Let
\beq
W:= \bw^3_\KK \KK^6 \, , \quad  \dim_\QQ W = 40 = 2 \cdot \dim_\KK W\, .
\eeq
The representation $h$ induces a representation $h_3$ on $W \otimes_\QQ \bRR$,
\beq
h_3 := \bw^3 h : \bCC^*\larr GL(W\otimes_\QQ \bRR) \, ,
\eeq
defined by
\beq
h_3(z) \left( u \wek v \wek w\right) :=  h(z) u \wek h(z) v \wek h(z) w \, .
\eeq
In the same way, tensorizing the $\KK$-vector space $W$ with $\bRR$, we obtain a complex vector space with a natural decomposition:
\beq
W \otimes_\QQ \bRR = \bigoplus_{p+q=3} W^{p,q} \, ,
\eeq
with
\beq
W^{p,q} = \{w \in W \otimes_\QQ \bRR  \, | \, h_3 (a + b i) w = (a + bi)^p (a-bi)^q w \} \, .
\eeq
Note that
$$W\otimes_\QQ \bRR =  (\bw^3_\KK \KK^6)\otimes_\QQ \bRR \simeq \bw^3_\KK \bCC^6 = \bCC^{20} \, ,$$
which is indeed the three-antisymmetric representation used in \cite{FFS}.\\
A rational Hodge structure of weight $k (\in\bZZ)$ is a $\QQ$-vector space $V$ with a decomposition of its complexification $V_\bCC:=V\otimes_\QQ\bCC$:
\beq
V_\bCC=\bigoplus\limits_{p+q=k}V^{p,q},\qquad{\rm and}\quad \ov{V^{p,q}}=V^{q,p}
\qquad(p,\,q\in\bZZ).
\eeq
We will prove the following result:
\begin{prop} There exists a $\QQ$-vector space $W_+ \subset W =  \bw^3 \KK^6$ of $\dim_\QQ 20$ such that
\beq
W_{+,\bCC}:= W_+ \otimes_\QQ \bCC = \bCC^{20} \quad \mbox{and } \quad W_+^{p,q} = \ov{W_+^{q,p}}\, ,
\eeq
i.e.\ $W_+$ carries a rational Hodge structure. \end{prop}
\noindent
This amounts to find a vector space $W_+$ such that the representation $h_3$ leaves $W_+ \otimes_\QQ \bRR$
invariant.

\subsection{The $\KK$-antilinear automorphism}
\label{sec:automorphism}
The existence of a space $W_+$, admitting a rational Hodge structure, is related to the existence of a natural $\KK$-antilinear automorphism $t$ of
$\bw_\KK ^3 \KK^6$ such that $t \circ t = \rm{const} \cdot \rm{Id}$.
The eigenspaces of $t$ generate a decomposition of $W=\bw_\KK^3 \KK^6$ and will define (isomorphic) Hodge structures. The automorphism $t$ is
the composition of two maps $\tau$ and $\rho$ which we now describe. 
\\

\noindent
For the first morphism consider the hermitian form $H$ on $\KK^6$ as a map
\beq H: \KK^6 \times \KK^6  \larr  \KK
\eeq
defined by
\beq
\label{acca}
H(v,w)= \sum_j \epsilon_j \ov{v_j} w_j
\eeq
with $\epsilon_j = +1$ for $j=1,2,3$ and $-1$ for $j=4,5,6$. So on the standard basis $\{e_1,e_2, \dots, e_6\}$ of $\KK^6$, $H$ is given by
\beq
H(e_i,e_j) = \epsilon_i \delta_{ij}.
\eeq
$H$ can be extended to a form $\tild{H}$ on $\bw\limits^3 \KK^6$. To this end we first define $\tild{H}$ on $(\KK^6)^3$ and then show that it is alternating and
$\KK$-linear in the second variable and $\KK$-antilinear in the first variable. Define
\beq
\begin{array}{rrcl}
\tild{H}: & (\KK^6)^3 \times  (\KK^6)^3 & \larr & \KK  \, , \\
& (\alpha, \beta) & \lmt & \tild{H}(\alpha, \beta)
\end{array}
\eeq
by
\beq
 \tild{H}((a, b, c), (p, q, r))  = \det\left[
 \begin{array}{ccc}
 H(a,p) & H(a,q) & H(a,r) \\
 H(b,p) & H(b,q) & H(b,r) \\
 H(c,p) & H(c,q) & H(c,r)
 \end{array}
 \right].
\eeq
Then $\tild{H}$ is separately alternating in the first and in the second factor.
Moreover, since $H$ itself is $\KK$-linear in the second variable and $\KK$-antilinear in the first, so is $\tild{H}$, thus
defining a map (to which we give the same name)
\beq
\begin{array}{rccc}
\tild{H}: & \bw\limits_{\KK }^3 \KK^6 \times  \bw\limits_{\KK }^3 \KK^6 & \larr & \KK  \, , \\
& (\alpha, \beta) & \lmt & \tild{H}(\alpha, \beta).
\end{array}
\eeq
by
\beq
 \tild{H}( a\wek b\wek c, p\wek q\wek r)  =\tild{H}((a, b, c), (p, q, r)).
\eeq

\noindent
Since $\tild{H}$ is $\KK$-linear in the second variable and $\KK$-antilinear in the first one,
it induces a $\KK$-antilinear map
\beq
\begin{array}{rrcl}
\tau: & \bw\limits_{\KK }^3 \KK^6 & \larr &  \rm{Hom}_\KK \left(  \bw\limits_{\KK }^3 \KK^6, \KK\right) \\
& \alpha & \lmt & \left[\beta \lmt \tild{H} (\alpha,\beta)\right] \, .
\end{array}
\eeq
Evaluating $\tild{H}$ on the basis of $ \bw\limits_{\KK }^3 \KK^6$, $\{e_i \wek e_j \wek e_k \}$ , $1\leq i<j<k \leq 6$, we get
$$
\begin{array}{r}
\tild{H} (e_i \wek e_j \wek e_k, e_i \wek e_j \wek e_k) =\epsilon_i \epsilon_j \epsilon_k  = \left\{\begin{array}{rcl}
1& \mbox{for} & \{i,j,k\} = \{1,2,3\} \\
-1 & \mbox{for} & \{i,j\} \subseteq \{1,2,3\} \, , k \in \{4,5,6\} \\
1& \mbox{for} & i \in \{1,2,3\} \, , \{j,k\} \subseteq \{4,5,6\} \\
-1 & \mbox{for} & \{i,j,k\} = \{4,5,6\}
\end{array}
\right.
\end{array}
$$
so that $\tau$ acts as
\beq
e_i \wek e_j \wek e_k \stackrel{\tau}{\lmt} \epsilon_i \epsilon_j \epsilon_k (e_i \wek e_j \wek e_k )^* \, ,
\eeq
where $(e_j \wek e_k \wek e_l )^*$ is the dual basis of $e_j \wek e_k \wek e_l$.\\

\noindent
The second morphism is induced by the isomorphism
$$
\begin{array}{rcccl}
\gamma: &\bw\limits_{\KK }^3 \KK^6 \times \bw\limits_{\KK }^3 \KK^6  & \stackrel{\simeq}{\larr} & \bw\limits_{\KK }^6 \KK^6  & \stackrel{\simeq}{\larr}  \KK  \\
&(\theta,\eta) &\lmt & \theta \wek \eta \, , &
\end{array}
$$
where the last map is the isomorphism sending $e_1\wek e_2 \wek e_3\wek e_4 \wek e_5\wek e_6$ to $1$. \\
Hence we get an isomorphism
$$
\begin{array}{rrcl}
\rho: & \bw\limits_\KK ^3 \KK^6 &\stackrel{\simeq}{\larr}& \rm{Hom}_{\KK }\left( \bw\limits_\KK^3 \KK^6, \KK \right) \\
& \alpha & \lmt &[ \beta  \lmt  \gamma(\alpha \wek \beta) ] \, ,
\end{array}
$$
which acts as
\beq
e_i \wek e_j \wek e_k  \stackrel{\rho}{\lmt} \delta_{ijk} (e_l \wek e_m \wek e_n )^*
\eeq
where $l,m,n \in \{1, \dots, 6\} \backslash \{i,j,k\}$ and $\delta_{ijk} =  \pm 1$ are suitable signs specified in appendix \ref{app:Hodge}.\\

\noindent
Using $\tau$ and $\rho$, we can define the automorphism $t$
\beq
t :=  \rho^{-1} \circ \tau: \bw\limits_\KK ^3 \KK^6  \larr  \bw\limits_\KK ^3 \KK^6.
\eeq
Obviously, since $\tau(w) = \rho(t(w))$, we have
\beq
\label{hgt}
\tild{H}(v,w)= \gamma (t(w) \wek v) \, .
\eeq
Since $\rho$ is $\KK$-linear and $\tau$ $\KK$-antilinear, it follows that $t$ is $\KK$-antilinear. We can write the action of $t$ explicitly on the
elements of a basis of $ \bw_\KK^3 \KK^6$.
This is shown in appendix \ref{app:Hodge}. In this way one easily verifies that $t^2= \rm{Id}$, and that $t$ has the eigenvalues $\pm1$, each with
multiplicity $10$. Thus we get the decomposition

\beq
W \simeq W_+ \oplus W_- \, , \quad \quad \dim_\QQ W_+ =\dim_\QQ W_- =20\, .
\eeq
\\
Note that $W_\pm$ are not $\KK$-vector spaces, since the automorphism $t$ is $\KK$-antilinear.

\subsection{Hodge structure on $W_+$}\noindent
Now we will prove that $W_\pm \subset W$ are Hodge structures.
\noindent
Since $( \bw_\KK^6 \KK^6 ) \otimes \bRR \simeq \bw^6 \bCC^6 \simeq \bCC$, we have
$$ h_6(z) (\alpha \wek \beta) := h_3(z)\alpha \wek h_3(z) \beta = |z|^6 (\alpha \wek \beta) \quad \mbox{for all } \alpha,\beta \in W.
$$
On the other hand, since $H(J v, J w) = H(v,w)$ (which implies $H(v,Jw)=H(Jv,J^2w)= -H(Jv,w)$)  and $h(a+b i)v=(aI +b J)v$, we also get
$$
\begin{array}{rcl}
H(h(a+b i)v,h(a+b i)w) = H((aI +b J)v, (aI +b J)w) =  (a^2+b^2) H(v,w) \, .
\end{array}
$$
By definition of $\tild{H}$
$$\tild{H}(h_3(z)\alpha, h_3(z) \beta) = (z \ov{z})^3 \tild{H}( \alpha,\beta ) =  |z|^6 \tild{H}( \alpha ,\beta )  \, .$$
Hence, using \eqref{hgt}, we conclude that
\beq
\begin{array}{rcl}
\gamma(t(h_3(z)\alpha) \wek h_3(z) \beta) &=& \tild{H}(h_3(z)\alpha, h_3(z) \beta) =  |z|^6 \tild{H}( \alpha,\beta )  = |z|^6  \gamma(t(\alpha) \wek \beta) \\
&=&  \gamma(h_3(z) t(\alpha) \wek h_3(z) \beta) \, ,
\end{array}
\eeq
for all $v,\beta \in W$, where in the last equality we used the fact that $\gamma$ is $\KK$-bilinear and, once we take its $\bRR$-linear extension, it
becomes $\bCC$-linear. This shows that $t \circ h_3(z) = h_3(z) \circ t$, i.e.\ $t$ preserves the decomposition given by $h_3$. \\

\noindent
Now, from $t^2 = \rm{Id}$, we obtain a decomposition
\beq
W \otimes_\QQ \bRR = (W_+  \otimes_\QQ \bRR) \oplus (W_- \otimes_\QQ \bRR),
\eeq
where
\beq W_\pm = \ker(t \pm \rm{Id}).
\eeq
Correspondingly the representation $h_3$ decomposes as
\beq
h_3(z)= \begin{pmatrix} h_3^+(z) & 0 \\
0 & h_3^-(z) \end{pmatrix} \, ,
\eeq
where $h_3^\pm (z) : W_\pm \otimes_\QQ \bRR \ra W_\pm \otimes_\QQ \bRR$ is just $h_3(z)$ restricted to $W_\pm \otimes_\QQ \bRR$.
Moreover, it is easy to check that
\beq
\Delta:= (2 \Theta + I) : W_+ \stackrel{\simeq}{\larr} W_- \, ,
\eeq
where $\Theta$ is the operator of multiplication by $\omega$, is an isomorphism of Hodge structures. Hence
$$\bw^3 \KK^6 = W_+ \oplus W_- \simeq W_+ \oplus \Delta \cdot W_+.$$
Obviously, since $\Delta \in \KK$,
$$W_+ \oplus \Delta \cdot W_+ \subseteq  W_+ \otimes_\QQ \KK.$$
For dimensional reasons the equality holds.\\
Since $\bw^3 \KK^6 = W_+ \otimes_\QQ \KK$ and $W_+ \otimes_\QQ \bCC = (W_+ \otimes_\QQ \KK ) \otimes_\KK \bRR,$we get
\beq
\big( \bw^3 \KK^6 \big) \otimes_\QQ \bRR \simeq W_+ \otimes \bCC.
\eeq
The Hodge structure of $W_+$ is now completely determined by
\beq
(W_{+,\bCC})^{3,0} = \bw^3 V^{1,0} \, ,  \quad (W_{+,\bCC})^{2,1} = \bw^2 V^{1,0} \otimes V^{0,1}\, ,
\eeq
and analogous expressions for $(W_{+,\bCC})^{1,2}$ and $(W_{+,\bCC})^{0,3}$.\\

\noindent
\section{Variation of Hodge structures and Special geometry}\label{sec:special}
\noindent
The original formulation of Special Geometry arose in the context of N = 2 supersymmetric theories coupled to supergravity and relied on the existence
of a holomorphic prepotential function $\cF$ (\cite{Stro}, \cite{CRTP}). \\
A \emph{projective special K\"ahler structure} \cite{Fr} is a special type of variation of polarized Hodge structures of weight 3 with Hodge numbers $h^{3,0}=1$
and $h^{2,1}=n$, a so-called variation of Hodge structures of Calabi-Yau type (\cite{BG}, \cite{CGG}). \\
To define a projective special K\"ahler structure on an $n$-dimensional K\"ahler manifold $(M,  \omega)$, together with an holomorphic line bundle
$L\ra M$,  one needs a holomorphic vector bundle $V \ra M$  of rank $n + 1$ with a given holomorphic inclusion $L \hookrightarrow V$ and a
flat connection $\nabla$ on the underlying real bundle $V_\bRR \ra M$, such that $\nabla(L) \subset V$ and the section
$$
\begin{array}{rcl}
M & \larr & \PP\left[V_\bRR \otimes \bCC \right] \\
m & \lmt &L_m
\end{array}$$
is an immersion. Finally one requires a nondegenerate alternating form $Q$ on $V_\bRR$ which has type (1,1) with respect to the
complex structure and satisfies $\nabla Q = 0$.

\

\noindent
For a CY threefold $X$, only the Hodge structure on the third cohomology group is of interest, as $H^2 (X, \bCC) = H^{1,1} (X )$. The Hodge structure on $H^3 (X , \bZZ)$ is the decomposition of its complexification:
$$H^3 (X, \bCC) = H^{3,0}(X ) \oplus H^{2,1}(X ) \oplus H^{1,2}(X ) \oplus H^{0,3}(X ) \, , \quad \ov{H^{p,q}} (X ) = H^{q,p} (X ) \, .$$
The intersection form on $H^3 (X, \bZZ)$ defines a polarization $Q_X$ on this Hodge structure. The polarization $Q_X$ is a symplectic form (so it is
non-degenerate, unimodular and alternating) and extends to a Hermitian form $H_X := iQ_X$ on $H^3 (X, \bCC)$ for which the Hodge decomposition is orthogonal:
$$H_X (v, w) = 0 \quad \mbox{if} \quad v \in H^{p,q}(X ), \, w \in H^{r,s}(X ) \quad \mbox{and} \quad (p, q) \neq (r, s) \, .$$
Moreover  $H_X$ is positive (negative) definite on $H^{3,0}(X )$ and $H^{1,2}(X )$ (on $H^{2,1}(X )$ and $H^{0,3}(X )$).\\
By the Bogomolov-Tian-Todorov Theorem (\cite{B}, \cite{Ti}, \cite{To}) the deformations of CY manifolds are unobstructed.
Hence there is a neighborhood $B$ of $0 \in H^1(X, T_X )$ and a family
of CY threefolds $\pi: \cX \ra B$ with fiber $\pi^{-1}(0) = X$ such that the period map $\cP : B \ra \cD$ has an injective differential.

\

\noindent
Let $\{\alpha_I (b),  \beta_J (b)\}$, $I,J=0, \dots n$,  be the dual basis to a symplectic basis of $H^3 (X_b , \bZZ)/\mathrm{torsion}$. Then we get, for
each $b \in B$, isomorphisms
\beq
H^3 (X_b , \bCC) \stackrel{\simeq}{\larr} \bCC^{2+2n} \, , \quad   \eta \lmt (\int_{\alpha_0} \eta, \dots, \int_{\alpha_n} \eta, \int_{\beta_0} \eta,
\dots, \int_{\beta_n} \eta) \, .
\eeq
This provides us with coordinates $t_0 , \dots  , t_n , s_0 , \dots  , s_n$ on each $H^3(X_b , \bCC)$. The symplectic form $Q$  in these coordinates is
then the two form
\beq Q  = dt_0 \wedge ds_0 + \dots  + dt_q \wedge ds_q \, .
\eeq
Since the period map has injective differential, we get a local isomorphism
\beq
\bCC \times B \larr \bCC^ {n+1} \, , \quad (s, b) \lmt \left(\int_{\alpha_0} s\Omega(b), \dots  , \int_{\alpha_n} s\Omega(b) \right) \, ,
\eeq
where $\Omega(b)$ spans $H^{3,0} (X_b )$ and its derivatives span $H^{3,0}(X_b) \oplus H^{2,1}(X_b)$.
There are holomorphic functions $F_0 , \dots  , F_n$ on $\bCC \times B$ which provide the other $n + 1$ coordinates:
\beq
F_j : \bCC \times B \larr \bCC, \quad F_j (t, b) := \int_{\beta_ j} s\Omega(b) \, .
\eeq
The restriction of $Q$  to $H^{3,0}(X_b) \oplus H^{2,1}(X_b)$ will be identically zero for all $b \in B$:
\beq
0 = dt_0 \wedge dF_0 + \dots  + dt_q \wedge dF_q \, ,
\eeq
As $dF_i = \sum_{j}  \partial F_i /\partial t_j \, \rmd  t_j$ this is equivalent to:
\beq
0 = \sum_{i,j} \frac{\partial F_i}{\partial t_j} dt_i \wedge dt_j \, , \quad \mbox{hence} \quad  \frac{\partial F_i}{\partial t_j} =
\frac{\partial F_j}{\partial t_i}
\eeq
for all $i, j$. Thus there exists a  \emph{prepotential} $\cF$ on $\bCC^{n+1}$, which satisfies $F_i = \partial \cF / \partial t_i$.\\
To the polarization $Q$ one can associate a cubic form $\Xi$ on the vector fields over $B$. Let $\xi_i\in T_pB$, $i=1,2,3$ three vector fields.
Then one defines
\begin{eqnarray}
\Xi: & T_pB \times T_pB \times T_pB & \longrightarrow \mathbb{C}, \cr
& (\xi_1, \xi_2, \xi_3) & \longmapsto Q(\Omega, (\xi_1 \xi_2 \xi_3 \Omega)).
\end{eqnarray}
The components of this cubic form define the Yukawa couplings.

\

\noindent
Let $\pi : \cX \ra M$ be a family of CY threefolds over a complex manifold $M$. We assume that $\dim M = \dim H^{2,1}(X_m) = n$ for all $m \in M$,
where $X_m := \pi^{-1}(m)$ is a CY threefold. Finally we require that the period map, which is well defined locally on $M$, has an injective differential
at all $m \in M$. \\
Hence we can define a fiber bundle $R^3 \pi_\ast \bZZ$, whose fiber is given by $H^3 (X_m, \bZZ)$ over $m \in M$, which is locally trivial.
The polarization $Q_m$ on $H^3 (X_m, \bZZ)$/torsion $\simeq \bZZ^{2q+2}$ is a nondegenerate alternating form. \\
We define a real vector bundle $V \ra M$ of rank $2n + 2$ on $M$, with a section $\omega$ of $\wedge^2 V^\ast$ as follows:
$$V := (R^3 \pi_\ast \bZZ) \otimes_\bZZ \bRR \, , \quad \omega_m := Q_m : V_m  \times V_m \larr \bRR \, ,$$
where we extended the polarization $\bRR$-bilinearly to the whole $V_m$, and $\omega_m$ is a symplectic form on $V_m$. Moreover, the vector bundle $V$ has a
flat connection $\nabla$ defined by imposing that the sections of $R^3 \pi_\ast \bZZ$ are flat. \\
From the polarization $Q_m$ one can derive the K\"ahler potential, whereas the cubic form $\Xi$ gives the Yukawa coupling.\\

\

\noindent
To study the special K\"ahler geometry of $\cM_{3,3}$, we perform an abstract variation of Hodge structures (VHS), which allows us to write down the prepotential
and the cubic form. \\
First we study the easiest example of VHS of CY type, namely $(1,1,1,1)$, to explicit the details of the construction and to show how
special geometry arises. Then we go over to the HS of type $(1,9,9,1)$, where we exhibit the special geometry structure of $SU(3,3)/S(U(3)\times U(3))$
and compare it with known results. A geometrical approach to derive this Hodge structure is provided in the appendix.

\subsection{VHS of type \boldmath{$(1,1,1,1)$}}
We consider the usual weight one variation of polarized Hodge structures on $(W_\bZZ := \bZZ e
\oplus \bZZ f , Q_W )$,
given by
\beq
W^{1,0}_{\bCC,\tau} := \langle \tau e + f \rangle, \quad W^{0,1}_{\bCC,\tau} :=\langle \ov\tau e + f \rangle, \quad Q_W (e, f ) = 1
\eeq
over the upperhalf plane $\mathbb {H} := \{ \tau \in \bCC : \rm{im}\, \tau > 0 \} \, .$\\
We can define a VHS of CY-type over $\HH$ by
\beq
(V_\bZZ, Q) := (\mathrm{Sym}^3 (W_\bZZ), \mathrm{Sym}^3 (Q_W ) ) \, .
\eeq
Thus the Hodge structure over $\tau \in \HH$ is:
\beq
\begin{array}{l}
V^{3,0}_{\bCC,\tau} := \langle(\tau e + f )^3 = \tau^3 e^3 + 3\tau^2 e^2 f + 3\tau ef^2 + f^3 \rangle , \\
V^{2,1}_{\bCC,\tau} :=\langle(\tau e + f )^2 (\ov\tau e + f ) = \tau^2 \ov\tau e^3 + (\tau^2 + 2\tau \ov\tau )e^2 f + (2\tau + \ov\tau )ef^2 + f^3 \rangle ,
\end{array}
\eeq
and
\beq
V^{p,q}_{\bCC,\tau} = \ov{V^{q,p}_{\bCC,\tau} }.
\eeq
Note that $\dim V_\bCC = \dim V_\bZZ \otimes_\bZZ \bCC = 4$ and the Hodge structure is of type $(1,1,1,1)$. \\
The polarization $Q : V_\bZZ \times V_\bZZ \ra \bZZ $ is the unique alternating form which is invariant for the action of $SL(2)$ on $V = \mathrm{Sym}^3 (W )$.
Its nonvanishing symplectic pairings are
\beq
Q(e^3 , f^3 ) = -Q(f^3, e^3 ) = 3, \quad Q(e^2f , ef^2 ) = -Q(ef^2, e^2 f ) = -1.
\eeq
Indeed, note that $V^{3,0}_{\bCC,\tau} \oplus V^{ 2,1}_{ \bCC,\tau} $ is isotropic for $Q$.

\noindent
Let
\beq
\Omega_{3,0}(\tau ) := \tau^3 e^3 + 3\tau^2 e^2 f + 3\tau e f^2 + f^3 \quad ( \in V^{3,0}_{\bCC,\tau} ),
\eeq
the generator of $V^{3,0}_{\bCC,\tau}$.
The cubic form $\Xi$ is also easily determined. Let us define the vector fields $\xi_i :=\lambda_i \partial_\tau \in T_\tau \HH$.
Since $\partial^3_\tau \Omega =6 e^3$, we get
\beq
Q(\Omega(\tau ), (\xi_1 \xi_2 \xi_3 \Omega)(\tau )) = Q(\tau^3 e^3 + 3\tau^2 e^2 f + 3\tau e f^2 + f^3 , 6\lambda_1\lambda_2\lambda_3 e^3 ) = 18\lambda_1\lambda_2\lambda_3 \, .
\eeq
Let $v= (F_0,F_1,t_1,t_0) =(\tau^3 s , 3\tau^2 s, 3\tau s, s)$, $s \in \bCC$. Hence $s = t_0$, $\tau = \frac{t_1}{3 t_0}$ and
$$F_0 = \frac{t_1^3}{27 t_0^2} \, , \quad F_1 = \frac{t_1^2}{3 t_0}  \, .$$
Then $Q$ can be written as:
$$
Q = 3 \rmd F_0 \wedge \rmd t_0 - \rmd F_1 \wedge \rmd t_1 \, .
$$
In order to derive the prepotential we have to change to a symplectic basis in which $\ov{t}_0 = 3 t_0 \, , \ov{t}_1 = -t_1 $. Then 
$$ F_0 = - \frac{\ov{t}_1^3}{3 {t}_0^2} \, , \quad F_1 = \frac{\ov{t}_1^2}{\ov{t}_0}  \, ,$$ 
and 
\beq
Q =  \rmd F_0 \wedge \rmd \ov{t}_0 + \rmd F_1 \wedge \rmd \ov{t}_1 = 0\, .
\eeq
Therefore
\beq
\frac{\partial F_0}{\partial \ov{t}_1}= - \frac{\ov{t}_1^2}{ \ov{t}_0^2} =\frac{\partial F_1}{\partial \ov{t}_0},
\eeq
and the prepotential $\cF$ is given by
$$ \cF= \frac{\ov{t}_1^3}{3 \ov{t}_0}.$$

\subsection{VHS of type \boldmath{$(1,9,9,1)$}}
\noindent
We now consider the variation of the Hodge structure on $W_+$. We can parameterize $SU(3,3)/S(U(3) \times U(3))$ with the entries $z_{ij}$ of the matrix
$Z \in  \cS_M = \{Z \in M(3, \bCC) \, : \,  I - {}^t \ov{Z} Z  > 0\} $. Hence we can construct a VHS over $\cS_M$:
\beq
\begin{array}{llllll}
V_\bZZ & \subset & \VV & \subseteq & \cS_M \times \big( \bw^3 \KK^6  \big) \otimes \bRR\\
\downarrow & &\downarrow \\
Z & \in & \cS_M
\end{array}
\eeq
Obviously
\beq
H_{|_{V^+ \times V^+} }> 0 \quad \mbox{and } \quad H_{|_{V^- \times V^-}} < 0 \, .
\eeq
Let $v_\pm \in V_\pm$. Then, by the definition of $V_\pm$ \eqref{eigensp},
\beq
\begin{array}{rcl}
i H(v_+,v_-) &=& H(i v_+,v_-) = H(J v_+, v_-) \stackrel{J \in U(H)}{=} H(J^2 v_+, J v_-) \\
&=& - H(v_+, -i v_-) =- \ov{(-i)} H(v_+,v_-) = -i H(v_+,v_-) \, ,
\end{array}
\eeq
$H$ being $\bCC$-(anti)linear in the first (second) variable. Hence $H(v_+,v_-)=0$ for every $v_\pm \in V_\pm$, i.e. $V_+$ and $V_-$ are orthogonal with respect
to $H$. Thus, if $(V_+)_Z = \begin{pmatrix} Z \\ I \end{pmatrix}$, then $(V_-)_Z = \begin{pmatrix} I \\ {}^t \ov{Z} \end{pmatrix}$.\\
\noindent
Let $e_1, \dots, e_6$ be a basis of $\bCC^6$. Then $(V_+)_Z$ is generated by $f_1,f_2,f_3$, whereas $(V_-)_Z$ is generated by $g_1,g_2,g_3$ with
\beq
\begin{array}{cc}
\begin{array}{rcl}
f_1&=& z_{11} e_1 + z_{21} e_2 + z_{31} e_3 + e_4 \, , \\
f_2&=& z_{12} e_1 + z_{22} e_2 + z_{32} e_3 + e_5 \, , \\
f_3&=& z_{13} e_1 + z_{23} e_2 + z_{33} e_3 + e_6 \, .
\end{array} &
\begin{array}{rcl}
g_1&=& e_1 + \ov{z_{11}} e_4 + \ov{z_{12}} e_5 + \ov{z_{13}} e_6  \, , \\
g_2&=& e_2 + \ov{z_{21}} e_4 + \ov{z_{22}} e_5 + \ov{z_{23}} e_6 \, , \\
g_3&=& e_3 + \ov{z_{31}} e_4 + \ov{z_{32}} e_5 + \ov{z_{33}} e_6 \, .
\end{array}
\end{array}
\eeq
Since
\beq
(W_{\bCC, Z})^{p,q} \simeq \bw^p (V_+)_Z \otimes \bw^q (V_-)_Z \, ,
\eeq
we obtain
\beq
\begin{array}{ll}
(W_{\bCC, Z})^{3,0} = \langle f_1 \wedge f_2 \wedge f_3 \rangle \, , & (W_{\bCC, Z})^{2,1} = \langle f_i \wedge f_j \wedge g_k \rangle \quad (i<j) \, ,\\
(W_{\bCC, Z})^{1,2} = \langle f_i \wedge g_k \wedge g_l \rangle \quad  (k<l)\, , &(W_{\bCC, Z})^{0,3} = \langle g_1 \wedge g_2 \wedge g_3 \rangle \, ,
\end{array}
\eeq
for $i,j,k,l \in \{2,3\}$.
Hence
\beq
\begin{array}{rcl}
(W_{\bCC, Z})^{3,0} &=&  \langle \Omega_{3,0}(Z) \rangle= \langle f_1 \wedge f_2 \wedge f_3 \rangle \\
 &=&  \det Z \, e_1\wedge e_2\wedge e_3 \, +  \sum\limits_{\substack{
i,j=1}}^3 \Big(  \sum\limits_{\substack{
k<l, \\
 k,l \neq i}} (-1)^{i+1} \frac{\partial}{\partial z_{ij} } \det Z \,  \, e_k\wedge e_l\wedge e_{j+3} \Big) \, +  \\
& & + \sum\limits_{\substack{
i,j=1}}^3  \Big( \sum\limits_{ \substack{
 k<l ,\\
 k,l \neq j+3}} (-1)^{j+1}  z_{ij} \, e_i\wedge e_k\wedge e_l \Big) \, + 
 e_4\wedge e_5\wedge e_6 
\end{array}
\eeq
Let $v= (F_{00},F_{11}, \dots, F_{33}, t_{00},t_{11} , \dots , t_{33})$, where $t_{00}=s$, $t_{ij}= s (-1)^{j+1} z_{ij}$,
\beq
F_{00}= s \det Z \quad \mbox{and} \quad F_{ij}  = s (-1)^{i+1} \frac{\partial}{\ z_{ij}} \det Z \, .
\eeq
Hence, in terms of the new coordinates $t_{ij}$, we get
\beq
F_{00}=  \frac{\partial}{\partial t_{00}} \Big( \frac{1}{t_{00}} \det T \Big) \quad \mbox{and} \quad F_{ij}  = (-1)^{i+j+1}  \frac{\partial}{\partial t_{ij}}  \Big( \frac{1}{t_{00}} \det T \Big)\, ,
\eeq
where $T=(t_{ij})_{i,j=1,2,3}$.\\

\noindent
The polarization $Q : V_\bZZ \times V_\bZZ \ra \bZZ $  is induced by the pairing given by the wedge product:
\beq
Q(e_i \wedge e_j \wedge e_k, e_l \wedge e_m \wedge e_n) = \pm 1 \quad \mbox{for }\quad \{i\dots n\}=\{1, \dots, 6\} \, .
\eeq
On the basis of $(W_{\bCC, Z})^{3,0}$ it takes the form
\beq
\begin{array}{rcl}
Q &=& \rmd F_{00} \wedge \rmd t_{00} -  \rmd F_{11} \wedge \rmd t_{11}  + \rmd F_{12} \wedge \rmd t_{12}  -  \rmd F_{13} \wedge \rmd t_{13} +  \rmd F_{21} \wedge \rmd t_{21} + \\
&& - \rmd F_{22} \wedge \rmd t_{22}  +  \rmd F_{23} \wedge \rmd t_{23}  -  \rmd F_{31} \wedge \rmd t_{31} + \rmd F_{32} \wedge \rmd t_{32} -  \rmd F_{33} \wedge \rmd t_{33} \, .
 \end{array}
\eeq
\noindent
Setting $\ov{t}_{00}=t_{00}$ and $\ov{ t}_{ij}=(-1)^{i+j+1}  t_{ij}$, we obtain
\beq
Q = \sum_{ij} \rmd F_{ij} \wedge \rmd \ov{t}_{ij} \, .
\eeq
Now $\frac{\partial F_{ij}}{\partial \ov{t}_{kl}} =  \frac{\partial F_{kl}}{\partial \ov{t}_{ij}} $. Hence there exists a prepotential $\cF$ given by
$$ \cF= -\frac{\det T}{\ov{t}_{00}} \quad \mbox{such that } \quad F_{ij} = \frac{\partial \cF}{\partial \ov{t}_{ij}} \, .$$

 \noindent
By definition, setting $\Omega_{3,0}(Z) \equiv \Omega$, the cubic form $\Xi$ is
\beq
\Xi : (\xi , \xi , \xi)\larr Q(\Omega , (\xi \xi \xi \Omega) \, .
\eeq
Take $\xi :=\sum_{i,j=1}^3 \lambda_{ij} \partial/\partial z_{ij} \in T_Z \cS_M$. Since
\beq
\renewcommand{\arraystretch}{2}{
 \begin{array}{rcl}
e_1\wedge e_2 \wedge e_3 &=&  \frac{\partial^3 \Omega }{\partial z_{11} \partial z_{22} \partial z_{33}}
=  \frac{\partial^3 \Omega }{\partial z_{12} \partial z_{23} \partial z_{31}}
=  \frac{\partial^3 \Omega }{\partial z_{13} \partial z_{21} \partial z_{32}}
=  -\frac{\partial^3 \Omega }{\partial z_{13} \partial z_{22} \partial z_{31}} \\
&=& - \frac{\partial^3 \Omega }{\partial z_{11} \partial z_{23} \partial z_{32}}  = -  \frac{\partial^3 \Omega }{\partial z_{12} \partial z_{21} \partial z_{33}}  \, ,
 \end{array}}
\eeq
and $0$ otherwise, we get:
\beq
Q \left(\Omega, (\xi \xi \xi \Omega)\right)= - 6 \det(\lambda_{ij}) \, .
\eeq

\

\section{Conclusions}
\noindent
In this paper we studied from an algebraic point of view the main properties of the cohomology structures of CY-type $(1,9,9,1)$ arising
as a mirror of the K\"ahler structures of the rigid orbifold $Z:=E^3/\bZZ_3$. We have realized it as the Hodge structures of weight $3$
over $\Lambda^3 (\mathbb{Q}[\omega])^6$, $\omega$ being a primitive third root of unity. The corresponding moduli space of Hodge structures resulted
to be just $\cM_{3,3}$, the Teichm\"uller covering of the K\"ahler moduli space of $Z$. We did not try to construct a family of varieties having
this cohomology substructure. The abstract algebraic construction of the Hodge structure results to be rich enough to determine the special K\"ahler geometry allowing to compute the K\"ahler potential and the holomorphic prepotential function.\\
The main advantage of the algebraic approach is to provide quite general results, independent from the details of the underlying geometrical
structures. However, it would be interesting to solve the problem of whether or not there exists a family of manifolds having the given Hodge structure
as third cohomology group, and what kind of geometrical properties such family should have. A hint in this direction is given in the Appendix, where we show how the $(1,9,9,1)$ Hodge structure arises as substructure of the third cohomology group of a complex six dimensional Abelian variety of Weil type. Thus, for example, one may expect to select the family of mirror varieties as embedded into this Abelian variety.
A detailed analysis of this geometrical characterization will be presented in a separated paper.

\

\noindent
\subsection*{Acknowledgements}
{\it The authors are greatly indebted to Bert van Geemen for helpful advices and stimulating discussions. We would also like to thank Alice Garbagnati for valuable comments and suggestions.}


\newpage
\appendix

\section{Where abelian varieties enter the game...} \label{app:abelian}
\noindent
We want to show that the Hodge structure of CY type $(1,9,9,1)$ can be recovered as polarized Hodge substructure $W_+$ of a bigger Hodge structure given by the third cohomolgy group of a six-dimensional abelian variety of Weil type.\\
First we show how the relation between $\cM_{3,3}=  SU(3,3)/S(U(3) \otimes U(3))$ and the moduli space of abelian varieties of Weil type arises. Then, choosing a specific lattice for an abelian variety of Weil type, we identify the Hodge structure of CY-type $(1,9,9,1)$ as Hodge substructure of the third cohomology group of this variety.
\subsection{Complex structures}
\label{sec:cpxstr}
\noindent
We can identify $V_\bRR= \bRR^{12} = \bCC^6 $ and view multiplication by $i$ as an $\bRR$-linear map on $\bRR^{12}$. So $i$ is a complex structure on the real vector space $\bRR^{12}$ and $\bCC^6 \simeq (\bRR^{12}, i)$.\\
Recall that the Hermitian form $H: \bCC^6 \times \bCC^6 \ra \bCC$ of signature $(3,3)$ is given by
\beq
H(v,w)= \sum_j \epsilon_j \ov{v_j} w_j
\eeq
with $\epsilon_j = +1$ for $j=1,2,3$ and $-1$ for $j=4,5,6$. Now for each
$$V_+ \in \cS_G=\{W \subset \bCC^6 : W \cong \bCC^3, H_{|_W} > 0 \} $$
we will define another complex structure $J = J_{V_+}$ on $\bRR^{12}$. \\

\noindent
We define $V_-:= V_+^\perp$, the perpendicular with respect to $H$ of $V_+$. Thus we obtain:
$$V_\bRR = V_+ \oplus V_- \, .$$
This decomposition defines a complex structure  on $V_\bRR$ (which corresponds to $V_+$) in the following way:
$$J= J_{V_+} :V_\bRR \larr V_\bRR \, , \quad \quad J v_+ = i v_+ \, , \quad J v_- = -i v_- \, ,$$
for every $v_\pm \in V_\pm$. By construction $J^2 = -I$ and thus we obtain a complex vector space $(V_\bRR, J)$.
$V_+$ and $V_-$ are the $+i$ and $-i$ eigenspaces of $J$ in $\bCC^6 = (\bRR^{12},i)$, so we can recover $V_+$ from the complex structure $J$.
Moreover, $J$ preserves the hermitian form $H$. Indeed, given the decomposition $V_\bRR = V_+ \oplus V_- $, we can write $v, w \in \bCC^6$ as
$v = v_+ + v_-$ and $w = w_+ + w_-$, then $J v = i v_+ - i v_-$ and analogously for $w$. Hence
$$\begin{array}{rcl}
H(Jv, Jw) &=& H(i v_+, i w_+) + H(i v_+, -i w_-) +H(-i v_-, i w_+) +H(-i v_-, -i w_-) \\
&=& H(v_+, w_+) + H(v_-, w_-) \\
&=& H(v, w) \, ,
\end{array}$$
where in the second equation we used the fact that, if $v_+ \in V_+$, then $i v_+ \in V_+$ (and analogously for $V_-$), and $V_+$ and $V_-$ are orthogonal with respect to $H$. Therefore $H(v_+,w_-) = H(v_-,w_+) = 0$.\\
Hence $J \in SU(H) = SU(3,3)$. We have an action of $SU(3,3)$ on the set of these $J$'s given by:
$$J \lmt J' : = A \, J \, A^{-1} \, , \quad A \in SU(3,3) \, .$$
If $V_+$ and $V_-$ are the eigenspaces of $J$, then the eigenspaces of $J'$ are given by $AV_+$ and $AV_-$. Indeed, by definition, $J' A = A J$, hence
$$\begin{array}{l}
J' A v_+ = A J v_+ = i A v_+\\
J' A v_- = A J v_- = - i A v_-\, ,
\end{array}$$
for $v_\pm \in V_\pm$. \\

\begin{remark} There is a natural way to embed $SU(3,3)$ in $Sp(20, \bCC)$. On a complex vector space $V$ of dimension $2n$, chosing a basis $\{e_i\}_{i=1, \dots, 2n}$, there exists a natural pairing:
$$\begin{array}{ccl}
\Phi: \bigwedge^{n} \bCC^{2n}  \times  \bigwedge^{n} \bCC^{2n} &\stackrel{\Phi}{\larr} & \bigwedge^{2n} \bCC^{2n} \simeq \bCC\\
(\omega, \theta) &\longmapsto& \omega \wedge \theta = \lambda e_1 \wedge \dots \wedge e_{2n} \longmapsto \lambda \, ,
\end{array}$$
where $\lambda \in \bCC$. So we put $\Phi(\omega, \theta) = \lambda$.\\
Let $A \in SL(2n, \bCC)$. We denote by $\tild{A}$ and $\doubletilde{A}$ the induced maps of $A$ on $\bigwedge^{n} \bCC^{2n}$ and  $\bigwedge^{2n} \bCC^{2n}\simeq \bCC$, respectively. Note that $\doubletilde{A}= \det \tild{A}$. Thus we have
$$(\tild{A} \omega) \wedge (\tild{A}\theta) = \doubletilde{A}(\omega \wedge \theta) = \det A \cdot (\omega \wedge \theta) = \omega \wedge \theta \, , $$
since $\det A = 1$. Thus $\Phi (\tild{A} \omega , \tild{A}\theta) =\Phi (\omega, \theta)$. \\
Focusing on the case $n=3$, we have $ \bigwedge^{3} \bCC^{6} \simeq \bCC^{20}$. The map $\Phi$ is alternating,
$$\Phi (\theta, \omega) = - \Phi (\omega, \theta) \, ,$$
and non-degenerate,
$$\Phi (e_i \wedge e_j \wedge e_k, e_l \wedge e_m \wedge e_n) = \pm 1 \quad \mbox{if $\{i, \dots, n\} = \{1, \dots, 6\}$} \, .$$
Hence $\Phi$ is a symplectic form and $\tild{A} \in Sp(\Phi) \simeq Sp(20, \bCC)$. Thus we get a homomorphism
$$\begin{array}{rcl}
 SL(6, \bCC)  \supset SU(3,3) & \larr & Sp(20, \bCC) \\
A &\lmt & \tild{A} \, .
\end{array}
$$
\end{remark}
\subsection{Hermitian form} (\cite{BL})
\noindent
The imaginary part of the Hermitian form $H$ is an alternating form $E: \bRR^{12} \times \bRR^{12} \ra \bRR$ given by $E(v,w):= \mathrm{Im} \, H(v,w)$. For $J=J_{|V_+}$ as in \ref{sec:cpxstr}, we define an $\bRR$-bilinear form $H_J: \bRR^{12} \times \bRR^{12} \ra \bCC$ by
\beq
H_J(v,w) = E(v,Jw) + i E(v,w) \, .
\eeq
\\
We show that $H_J$ is an Hermitian form on the complex vector space $(\bRR^{12}, J)$. Indeed,
\beq
\begin{array}{rcl}
H_J(v,Jw) &=& E(v,J^2 w) + i E(v,Jw) = - E(v,w) + i E(v, w)  \\
&=& i H_J(v,w)\, ,
\end{array}
\eeq
so $H_J$ is $\bCC$-linear (for $(\bRR^{12},J)$) in the second variable, and
\beq
\begin{array}{rcl}
H_J(w,v) &=&  E(w,Jv) + i E(w,v) = E(J w , J^2 v) + i E(w, v) \\
&=& - E(J w , v) + i E(w,v) = E(v,Jw) -i E (v,w)  \\
&=& \ov{H_J(v,w)}\, .
\end{array}
\eeq
Moreover, $H_J$ is positive definite. In fact,
\beq
\begin{array}{rcl}
H_J(v,v) &=& E(v,Jv) + i E(v,v) = E(v_+ + v_-,i v_+ -i v_-) + i E(v_+ + v_-,v_+ + v_-) \\
&=& E(v_+,i v_+) +E( v_-,i v_+ ) +E(v_+,-i v_-) +E(v_-,-i v_-) \, .
\end{array}
\eeq
Since $H$ has signature $(3,3)$ and is positive definite on $V_+$, it will be negative definite on $V_-$.  Thus $$
\begin{array}{lcl}
E(v_+,i v_+) = H(v_+,v_+) \geq 0 &\quad & E( v_-,i v_+ ) = \rR \rme \, H(v_-,i v_+) = 0  \\
E(v_+,-i v_-)  = \rR \rme \, H(v_+,i v_-) = 0  &\quad & E(v_-,-i v_-) = - H(v_-,v_-) \geq 0 \, .
\end{array}$$
If $v=v_+ + v_- \neq 0$, then $v_+$ or $v_- \neq 0$, hence
\beq
\label{R1}
E(v, Jv) = H_J(v,v) > 0
\eeq for all $0 \neq v \in \bRR^{12}$. \\

\noindent
Furthermore, since $H_J(Jv,Jw) = H_J(v,w)$,
\beq
\label{R2}
E(Jv,Jw) = E(v,w) \quad \mbox{for $v,w \in V_\bRR$} \, .
\eeq
Hence $E$ satisfies the Riemann conditions \eqref{R1} and \eqref{R2} for the complex vector space $(\bRR^{12},J)$. \\

\noindent Our goal is to construct abelian varieties of the form $X = (V_\bRR, J)/\La$ for a fixed lattice $\La \simeq \bZZ^{12}$. In order for $E$ to be a polarization on  the abelian variety $(\bRR^{12}, J)/\La$, all that remains to be done is to choose $\La$ in such a way that $E$ fulfills the integrality condition, i.e.\
\beq E(\lambda, \mu) \in \bZZ \, , \quad \mbox{for all } \lambda, \mu \in \La \, .
\eeq

\subsection{Lattice}
\noindent
We choose the lattice
\beq \La := (\bZZ[\omega])^6 \subseteq \bCC^6 \, , \quad \quad  \mbox{where } \omega^3 = 1 \, .\eeq
Note that multiplication by $i$ on $\bCC^6$ can be written in terms of the map $\Theta: \bCC^6 \ra \bCC^6$, which is given by $v\mapsto \omega v$, as
\beq
i = \frac{1}{\sqrt{3}} \left( 2 \Theta + \mathrm{Id} \right) =: \frac{1}{\sqrt{3}}  \Delta \, .
\eeq
The  alternating map $E': \bRR^{12} \times \bRR^{12} \ra \bRR$ given by
\beq \label{riemform} E' := \frac{2}{\sqrt{3}}\, \mathrm{Im}\, H : V_\bRR \times V_\bRR \larr \bRR \, ,
\eeq
is integer-valued on the lattice $\La$. Indeed, let $\lambda = (n_1+m_1\omega, \dots, n_6+m_6\omega)$, $\mu=(p_1+\omega q_1, \dots, p_6+q_6\omega) \in \La$,
then, setting $\epsilon_i = 1$ for $i=1,2,3$ and $-1$ for $i=4,5,6$,
\beq 
\label{Eprimo}
\begin{array}{rcl}
E'(\lambda, \mu)&=&  \frac{2}{\sqrt{3}} \mathrm{Im} H(\lambda, \mu) = \frac{2}{\sqrt{3}} \mathrm{Im}  \sum\limits_{i=1}^6 \epsilon_i \ov{(n_i+m_i\omega)} (p_i+\omega q_i)  \\
&=&  \sum\limits_{i=1}^6 \epsilon_i (n_i q_i - m_i p_i  ) \, .
\end{array}
\eeq
Obviously $E'$ still satisfies the Riemann conditions. Hence $X=(\bRR^{12}, J)/(\bZZ [\omega])^6$ with the polarization $E'$ is an abelian variety.
In particular, it is a principally polarized abelian variety, since the determinant of the alternating matrix defining $E'$ is equal to $1$.

\subsection{Abelian varieties of Weil type}
We will now prove that the abelian variety $X=(\bRR^{12}, J)/(\bZZ [\omega])^6$ has an additional structure, namely it is an abelian variety of Weil type.
\begin{definition}
An abelian variety of \emph{Weil-type} of dimension $2g$ is a pair $(X,\KK )$, where $X$ is a $2g$-dimensional abelian variety and $\KK  \hookrightarrow End(X) \otimes \QQ$ is an imaginary quadratic field such that for all $k \in \KK $ the endomorphism $t(k)$ has $g$ eigenvalues $k$ and $g$ eigenvalues $\ov{k}$:
$$
t(k) \mbox{ acts as} \quad \mathrm{diag} (\underbrace{k, \dots, k}_g, \, \underbrace{\ov{k}, \dots, \ov{k}}_g) \quad \mbox{on $T_0X$}
$$
\end{definition}
\noindent
(fixing an embedding $\KK  \subset \bCC$). Hence $t(k)^*$ has eigenvalues $k, \ov{k}$ with multiplicities $g$ on $H^{1,0}(X) = T_0X^*$.\\

\noindent
The abelian variety $X=(\bRR^{12}, J)/(\bZZ [\omega])^6$ obviously admits an automorphism of order three given by multiplication by $\omega$. We observe that
$$\omega \La = \La$$
and $\omega$ is also a $\bCC$-linear map for any $J=J_{|V_+}$ with $V_+ \in \cS_G$ it follows that:
$$\omega J = J \omega \, ,$$
since $i J = J i$. Hence $a + b\omega \in End(X)$, for all $a, b \in \bZZ$.\\
Now we see how the structure $(3,3)$ of Weil type arises. On the tangent space $T_0 (X)$ at the origin of $X$ we have the decomposition given by
$$(\bRR^{12},J) = T_0(X) = V_+ \oplus V_- \, .$$
Since $J$ acts as multiplication by $+i$ on $V_+$ and by $-i$ on $V_-$, the action of $\Theta$ is given by $\omega$ on $V_+$ and by $\ov{\omega}$ on $V_-$. Hence $a + b\omega$ acts as $(a+b\omega, a+b\omega, a+b\omega, \, a+b\ov{\omega}, a+b\ov{\omega} , a+b\ov{\omega}) \quad \mbox{on $T_0X$}$.\\
Thus $\KK:= \QQ(\sqrt{-3}) = \QQ(\omega)$ acts as
$$
\mathrm{diag} (\underbrace{\omega, \omega, \omega}_3, \, \underbrace{\ov{\omega},  \ov{\omega}, \ov{\omega}}_3) \quad \mbox{on $T_0X$} \, .
$$

\begin{example} Let $V_0 = \bCC e_1 \oplus \bCC e_2 \oplus \bCC e_3$ and $V_0^\perp = \bCC e_4 \oplus \bCC e_5 \oplus \bCC e_6$.  The complex structure $J_0$, associated to this decomposition, acts as $i$ on $V_0$ and as $-i$ on $V_0^\perp$. When we consider the quotient $X = \bCC^6/\bZZ[\omega]^6$, $J_0$ maps each of the $\bCC/\bZZ[\omega]$ into itself. Hence $X=E^6$, where $E=\bCC/\bZZ[\omega]$.
\end{example}

\noindent
To summarize, we proved that the choice of a $3$-dimensional subspace $W$ of a $6$-dimensional complex vector space $V\simeq \bCC^6$, such that there exists an Hermitian form $H$ on $V$, which is positive-definite on $W$, is equivalent to the choice of a complex structure $J$ on  the underlying real $12$-dimensional vector space, provided $J$ preserves $H$. Moreover, starting from $H$, one can define a Riemann form $E'$ on $(\bRR^{12},J)$. Choosing a lattice $\La$ on which $E'$ is integer-valued, the data $(V, \La, E')$  identify a polarized abelian variety $(X:=V/\La,E')$.
\begin{remark} In \cite{vG} the converse is proved: any $2n$-dimensional polarized abelian variety of Weil type is a member of an $n^2$-dimensional family of abelian varieties of Weil type parametrized by $SU(n,n)$.
\end{remark}

\section{Cohomology of Abelian varieties of Weil type} \label{app:cohom}
\noindent
Here we consider some technical points about the cohomology of the Abelian varieties of Weil type.

\subsection{From homology to cohomology.}\label{B1} Let $X= \bCC^6/\La$ be the abelian variety of complex dimension $g=6$ constructed in the previous section.
Then $\La \cong \pi_1 (X)$ and $\pi_1 (X) \cong H_1(X,\bZZ)$ (Hurewicz theorem), so we identify $\La = H_1(X,\bZZ) $. We also have $H_1(X, \bRR)  \simeq H_1(X,\bZZ) \otimes_\bZZ \bRR = \La \otimes_\bZZ \bRR \simeq \bCC^6$. \\
The dual of $H_1(X,\bZZ)$ is $$H^1(X,\bZZ) := \mathrm{Hom}_\bZZ (H_1(X,\bZZ), \bZZ)\, .$$
Let $E':  H_1(X, \bZZ)\times H_1(X, \bZZ) \ra  \bZZ$ be the Riemann form on $X$ defined by \eqref{riemform}. As $\det E' = 1$, it induces an isomorphism
$$
\begin{array}{rrcl}
D: & H_1(X,\bZZ) & \larr & H^1(X, \bZZ) \\
& v & \longmapsto & [w \longmapsto E'(v,w)]
\end{array}$$
which induces (after $\bRR$-linear extension) an isomorphism
$$D_\bRR : H_1(X,\bRR)  \stackrel{\cong}{\larr}  H^1(X, \bRR) \, .$$
The complex structure $J$ on $H_1(X,\bRR)$ induces, under the isomorphism $H_1(X,\bRR) \stackrel{\cong}{\larr} H^1(X, \bRR)$, a complex structure $D_\bRR J D_\bRR^{-1}$ on $H^1(X, \bRR)$, which we still denote by $J$.
\\
Multiplication by $i = \frac{1}{\sqrt{3}} \Delta$ gives $H^1(X, \bRR)$ the structure of a $6$-dimensional complex vector space: $\left(H^1(X, \bRR),i \right) \simeq \bCC^6$. By duality, there exists a decomposition of $H^1(X, \bRR)$:
\beq
\begin{array}{rrccc}
J \circlearrowright H^1(X, \bRR) &=& V^+ &\oplus & V^-\\
&&+i & & -i
\end{array}
\eeq
in terms of the dual vector spaces of $V_+$ and $V_-$.

\subsection{First cohomology group.} We proved that the complex vector space $H^1(X, \bCC) :=  H^1(X, \bRR) \otimes \bCC \simeq  \bCC^{12}$ decomposes under the action of $J$:
$$
\begin{array}{cccc}
H^1(X, \bCC) = & H^{1,0}(X) & \oplus & H^{0,1}(X) \\
& J = +i & & J=-i \, ,
\end{array}$$
where $H^{1,0}(X)$ and $H^{1,0}(X)$ are the (complex $6$-dimensional) eigenspaces of $J$ corresponding to the eigenvalues $+i$ and $-i$, respectively. \\

\noindent
Since $X= \bCC^{6}/\La$ is an abelian variety of Weil type of dimension $6$, with imaginary quadratic field $\KK  = \QQ(\om)= \{a + b \om : a,b \in \QQ\}$, where $\om^3 = 1$ and $\om \neq 1$, $X$ has an endomorphism $\Theta$ of order $3$, which induces a $\bZZ$-linear map $\Theta_* : \La \ra \La$ such that $\Theta_*^3 = \mathrm{Id}$. By definition of Weil type, the action of $\Theta^*$ on $H^1(X, \bCC)$ is represented by
$$diag(\underbrace{\om, \dots, \om}_{6}, \underbrace{\ov\om, \dots, \ov\om}_{6})\, .$$ \\

\noindent
Hence $\Theta^*$ provides another decomposition of $H^1(X, \bCC)$, namely the one into eigenspaces of $\Theta^*$:
$$
\begin{array}{cccc}
H^1(X, \bCC) \,  = & V & \oplus &\overline{V} \\
& \Theta^*=\om & & \Theta^* = \ov\om
\end{array}
$$
where $V$ and $\ov{V} $ are the ($6$-dimensional) eigenspaces of $\Theta$ corresponding to the eigenvalues $\om$ and $\ov\om$, respectively. Since the map $\Theta$ is holomorphic, one has $\Theta^* J = J \Theta^*$, which implies
$$\Theta^*(H^{1,0}) \subseteq  H^{1,0} \, , \qquad \Theta^*(H^{0,1}) \subseteq H^{0,1}  \, .$$
Thus we obtain the following refined decomposition of $H^1(X, \bCC)$ in eigenspaces of both $J$ and $\Theta^*$ :
$$\begin{array}{ccccccccccc}
H^1(X, \bCC) &=& V^{1,0} & \oplus  & \ov{V}^{1,0} & \oplus & V^{0,1}  &  \oplus  &\ov{V}^{0,1} \\
&=& \left( H^{1,0} \cap V \right) & \oplus  & \left( H^{1,0} \cap \ov{V}  \right) & \oplus & \left( H^{0,1} \cap V \right) &  \oplus  & \left( H^{0,1} \cap \ov{V}  \right) \\
&& J = +i & & J = +i & & J=-i & & J=-i\\
&& \Theta^*=\om & & \Theta^* = \ov\om & & \Theta^*=\om & & \Theta^* = \ov\om
\end{array}
$$\\
As we assume that $X$ is of Weil type, $V^{1,0} := H^{1,0} \cap V $ and $V^{0,1} :=  H^{0,1} \cap V $ have both dimension $3$ (and analogously for the subspaces $\ov{V}^{1,0}$ and $\ov{V}^{0,1}$ obtained by replacing $V$ with $\ov{V} $.) \\

\noindent
\subsection{Third cohomology group.} Now we focus on the third cohomology group of $X$. From the previous paragraph, $H^3(X, \bCC) = \bigwedge^3 H^1(X, \bCC) = \bigwedge^3 \left(H^{1,0}\oplus H^{0,1} \right)$. Thus the Hodge decomposition of $H^3(X, \bCC) $ is given by
\beq H^3(X, \bCC)  = \bigoplus_{p+q = 3} H^{p,q}(X) \quad \mbox{where} \quad H^{p,q}(X) := \bigwedge^p H^{1,0} \otimes \bigwedge^q H^{0,1} \, .
\eeq
Note that
\beq
H^{p,q} = \{w \in H^3(X, \bCC) \, | \, (a + b J) w = (a + bi)^p (a-bi)^q w \} \, .
\eeq

\noindent
The dimension of $H^3(X,\bCC)$ and $H^{p,q}(X)$ are displayed in the following table:

$$
\begin{array}{c|c|c|c|c}
H^3(X, \bCC) & H^{3,0} & H^{2,1} & H^{1,2} & H^{0,3} \\
\hline
\binom{12}{3} = 220 & \binom{6}{3}= 20 &  \binom{6}{2} \cdot 6= 90 &  6 \cdot \binom{6}{2} =90 &  \binom{6}{3}= 20
\end{array}
$$
Obviously, the map $\Theta^*: H^1(X, \bZZ) \ra H^1(X, \bZZ)$ induces a map $$\Theta_3^*: H^3(X, \bZZ) \larr H^3(X, \bZZ) \, .$$ The $\bCC$-linear extension to $H^3(X, \bCC)$ of  this map will still be denoted by $\Theta_3^*$.

\section{Hodge substructure of the third cohomology group}\label{app:Hodge}
\noindent
\subsection{The Hodge substructure $W$}
In Appendix \ref{app:cohom} we described how the cohomology of an abelian sixfold of Weil type admitting an endomorphism $\Theta$ of order $3$ decomposes. Now we want to prove that there exists a Hodge substructure $W$ of the third cohomology group on which $\Theta_3^*$, the map induced by $\Theta$, acts trivially. Furthermore we show that $W$ admits a Hodge substructure $W_+$ of type $(1,9,9,1)$. A similar construction for the second cohomology group of abelian fourfolds of Weil type was given in \cite{Lo}.\\

\noindent
Whilst $\Theta^*$ does not have $1$ as eigenvalue on $H^1(X, \bCC)$, ${\Theta_3}^*$ does have eigenvalue $1$ on $H^3(X, \bCC)$. We will use this to define a Hodge substructure of $H^3(X, \QQ)$ as follows:
$$W:= ker(\Theta_3^* - \mathrm{Id}) = \{w \in H^3(X, \QQ) \, | \, \Theta_3^*w = w\} \, .$$
Since both $\mathrm{Id}$ and $\Theta_3^*$ are morphisms of Hodge structures (so $\Theta_3^*(H^{p,q}) \subseteq H^{p,q}$, $p+q=3$), $W$ is a Hodge substructure, i.e.\
$$W \otimes \bCC = W^{3,0} \oplus W^{2,1} \oplus  W^{1,2} \oplus  W^{0,3} \qquad (\subseteq H^3(X, \bCC))\, ,$$
where
$$W^{p,q} := (W \otimes \bCC) \cap H^{p,q}(X)$$
satisfy
$$W^{q,p} = \ov{W^{p,q} } \, .$$
To find the dimensions of the subspaces $W^{p,q}$ and their relation to the $H^{p,q}$ we recall that $\Theta^*$ acts as $\om^a \, \ov\om^b$ on $\bigwedge^a V \otimes \bigwedge^b \ov{V}$. \\
Thus, as  $H^3(X, \bCC) = \bigoplus\limits_{a+b=3} (\bw\limits^a V ) \otimes (\bw\limits^b \ov{V} )$ and $\Theta^* = \mathrm{Id}$ on $W$, we have$$
W\otimes \bCC = \bigwedge^3 V \oplus \bigwedge^3 \ov{V}  \, .$$
Since $\dim_\bCC V = \dim_\bCC \ov{V}= 6$, $\dim (W\otimes \bCC) = \binom{6}{3} + \binom{6}{3} =40$. Hence $\dim_\QQ W =40$. \\
The Hodge decomposition of $W$ is given by
$$
\begin{array}{lccccl}
W^{3,0} & = &  (\bigwedge^3 V)^{3,0} & \oplus & (\bigwedge^3 \ov{V} )^{3,0} & \quad 2 = 1+1  \\
W^{2,1} & = &  (\bigwedge^3 V)^{2,1} & \oplus & (\bigwedge^3 \ov{V} )^{2,1} & \quad 18 =  9 + 9 \\
W^{1,2} & = &  (\bigwedge^3 V)^{1,2} & \oplus & (\bigwedge^3 \ov{V} )^{1,2} &  \quad 18 \\
W^{0,3} & = &  (\bigwedge^3 V)^{0,3} & \oplus & (\bigwedge^3 \ov{V} )^{0,3} & \quad 2
\end{array}
$$
where $V^{p,q} := V \cap H^{p,q} = \big(\bw^p V^{1,0} \big) \otimes \big( \bw^q V^{0,1} \big)$ and $\ov{V}^{p,q} := \ov{V} \cap H^{p,q} = \big(\bw^p \ov{V}^{1,0} \big) \otimes \big( \bw^q \ov{V}^{0,1} \big)$. \\
On the right of each subspace its dimension is displayed. Hence $W$ is a Hodge substructure of type $(2,18,18,2)$. We want to show that $W$ can be further decomposed into two Hodge substructures of CY type $(1,9,9,1)$.

\subsection{W as $\KK$-vector space}
Since $X$ is an abelian variety of Weil type with $\QQ(\om) \subset \rm{End}(X)$, we have an action of $\QQ(\om)$ on $H^1(X,\QQ)$. Recall that $\dim_\QQ H^1(X, \QQ) = 12$ and $\dim_{\QQ(\om)} H^1(X, \QQ) = 6$. \\

\noindent
Let $V= H^1(X, \QQ)$, $\KK= \QQ[\om]$. Hence $V$ is a $\KK$-vector space. If $e_1, \dots, e_6$ is a $\KK$-basis of $V$, then, taking $e_1, \dots, e_6, \Theta e_1, \dots, \Theta e_6$, we obtain a $\QQ$-basis of $V$. \\
Let $a, b, c \in V$. We define $\QQ$-multilinear maps
\beq
\begin{array}{c}
\eta_{ijk} :  V^3  \larr \bw\limits_\QQ^3 V \\
\eta_{ijk}(a,b,c) :=  (\Theta^*)^i a \wedge (\Theta^*)^j b  \wedge (\Theta^*)^k c \, .
\end{array}
\eeq
In particular, we denote $\eta(a,b,c) :=\eta_{000}(a,b,c) = a \wedge b \wedge c$. Then we get, using $\Theta^2 = -1 - \Theta$,
$$
\begin{array}{rcl}
\Theta_3^* \eta_{000} &=&  \eta_{111} \, ,\\
\Theta_3^* \eta_{111} & =:& \tild\eta \, ,
\end{array}
$$
where $\tild\eta := -\eta_{000} - \eta_{001} - \eta_{010} - \eta_{011} - \eta_{100} - \eta_{101} - \eta_{110} - \eta_{111}$. As $(\Theta_3^* )^3 = (\Theta_3^3)^*= \mathrm{Id}$, we get that $\eta_{000} +  \eta_{111} + \tild\eta$ is invariant under $\Theta_3^*$:
$$\Theta_3^* ( \eta_{000} +  \eta_{111} + \tild\eta)=  \eta_{000} +  \eta_{111} + \tild\eta \, ,$$
hence it lies in $W$. Thus we have a $\QQ$-multilinear map
\beq
\begin{array}{c}
w :   V^3 \larr W\\
w(a,b,c) := \eta_{000}(a,b,c) +  \eta_{111}(a,b,c) + \tild\eta(a,b,c) \, .
\end{array}
\eeq
Hence, for any $a,b,c \in V$, we have an element in $W$ of the form $w(a,b,c)$.
\\
We can now give a $\QQ$-basis for $W$. Taking the $\QQ$-basis $\{e_1, \dots, e_6, \Theta e_1, \dots, \Theta e_6\}$ of $V$, we consider the vectors of the form
\beq
w(e_i,e_j,e_k) \quad \mbox{and} \quad w(\Theta^* e_i,e_j,e_k) \, .
\eeq
These are $2 \cdot \binom{6}{3}= 2 \cdot 20 = 40$ vectors which belong to $W$. We show that they are linearly independent over $\QQ$. \\
In fact, by definition, $w(e_i,e_j,e_k)$ and $w(e_{i'},e_{j'},e_{k'})$ are linearly independent over $\QQ$ for $\{i,j,k\} \neq \{i',j',k'\}$, and analogously for the $w(\Theta^* e_i,e_j,e_k)$. We can write $w(\Theta^* e_i,e_j,e_k)$ explicitly in terms of the $\eta_{lmn}$:
$$\begin{array}{rcl}
w(\Theta^* e_i,e_j,e_k) &=&  \left(- \eta_{001} - \eta_{010} - \eta_{011} - \eta_{100} - \eta_{101} - \eta_{110}\right) \; (\Theta^* e_i,e_j,e_k) \\
&=&  ( \eta_{000} + \eta_{100} - \eta_{010} - \eta_{001} - \eta_{111}) \; (e_i,e_j,e_k) \\
&=&  \eta_{000}(e_i,e_j,e_k) + \dots  \, ,
\end{array}
$$
whereas $w(e_i,e_j,e_k)$ does not contain $\eta_{000}(e_i,e_j,e_k) = e_i\wedge e_j\wedge e_k$.  Thus also $w(e_i,e_j,e_k)$ and $w(\Theta^* e_i,e_j,e_k)$ are linearly independent over $\QQ$. Hence the vectors $w(e_i,e_j,e_k), w(\Theta e_i,e_j,e_k)$ form a $\QQ$-basis of the $40$-dimensional $\QQ$-vector space $W$. \\

\noindent
We will need the following result. Since $$w(a,b,c) = \Theta a \wedge b \wedge c +  a \wedge \Theta b \wedge c + a \wedge b \wedge \Theta c + \Theta a \wedge \Theta b \wedge c +\Theta a \wedge b \wedge \Theta c + a \wedge \Theta b \wedge \Theta c \, $$
we obtain that
$$
\begin{array}{rcl}
w(\Theta a, b, c) &=& \Theta^2 a \wedge b \wedge c +  \Theta a \wedge \Theta b \wedge c + \Theta a \wedge b \wedge \Theta c +  \Theta^2 a \wedge \Theta b \wedge c \, +\\
&& \Theta^2 a \wedge b \wedge \Theta c + \Theta a \wedge \Theta b \wedge \Theta c  \\
&=&   - a \wedge b \wedge c - \Theta a \wedge b \wedge c + \Theta a \wedge \Theta b \wedge c +  \Theta a \wedge b \wedge \Theta c -  a \wedge \Theta b \wedge c \, + \\
&& - \Theta  a \wedge \Theta b \wedge c  - a \wedge b \wedge \Theta c - \Theta a \wedge b \wedge \Theta c +  \Theta a \wedge \Theta b \wedge \Theta c\\
&=&  w(a, \Theta b, c) \, ,
\end{array}
$$
and analogously for $w(a, b, \Theta c)$. Hence
\beq
\label{wtetaabc}
w(\Theta a,b,c) = w( a,\Theta b,c) =w(a,b,\Theta c) \, .
\eeq
\\
\noindent
We want to give $W$ the structure of a $\KK$-vector space. For this we want to show that the map
$$\begin{array}{ccl}
W& \larr &W \hspace{2cm} (\subseteq \bw_\QQ^3 V)\\
a \wedge b \wedge c &\lmt & \Theta a \wedge b \wedge c
\end{array}$$
is well-defined. \\
First we observe that the map
$$\begin{array}{rccl}
\Phi : &V^3& \larr &W \\
& (a,b,c) &\lmt & w(\Theta a, b, c)
\end{array}$$
is $\QQ$-trilinear (since the $\eta$'s are $\QQ$-trilinear). Moreover, $\Phi$ is an alternating map: it is obviously alternating under $(b\leftrightarrow c)$ and, exchanging $a$ and $b$, gives:
$$\Phi(a,b,c) = w(\Theta a,b,c) = -w(\Theta b,a,c)=-\Phi(b,a,c) \, ,$$
since $w(\Theta a, b, c)=w(a, \Theta  b, c) = -w(\Theta b, a, c)$ using \eqref{wtetaabc}. Hence $\Phi$ induces a well-defined map $\tild\Phi : \bw_\QQ^3 V \larr W$. \\
Once we restrict $\tild\Phi$ to $W$, we obtain a map
$$\begin{array}{rccl}
\tild\Phi_{|W} = \Phi : &W& \larr &W \, ,\\
& w(a \wedge b \wedge c) &\lmt & w(\Theta a \wedge b \wedge c)
\end{array}$$
which is well-defined.
Thus $W$ has the structure of $\KK$-vector space, where $a+b \om \in \KK$ acts as $a + b \Phi$.\\
Since $\{w(e_i,e_j,e_k), w(\Theta e_i,e_j,e_k)\}$ is a $\QQ$-basis of $W$, we get that the $w(e_i,e_j,e_k)$ are a $\KK$-basis of $W$.
\\

\noindent
\subsection{Isomorphism with $\bw_\KK^3 H^1(X,\QQ)$}
Now we are able to prove the isomorphism between $W$ and $\bw_{\KK }^3 H^1(X,\QQ)$. Consider the map
\beq
\begin{array}{rrcl}
\Psi : & V^3 & \larr & W\\
& (a,b,c) &\lmt & w(a,b,c) \, .
\end{array}
\eeq
Then $\Psi$ is clearly $\QQ$-multilinear and alternating. We want to prove that $\Psi$ is also $\KK$-multilinear. In fact, by \eqref{wtetaabc}
$$
\begin{array}{l}
\Psi(\Theta a, b, c) = w(\Theta a, b, c) = \Theta w(a,b,c) \, ,\\
\Psi(a, \Theta b, c) = w(a, \Theta b, c) = \Theta w(a, b,c) \, ,\\
\Psi(a, b, \Theta  c) = w(a, b, \Theta  c) = \Theta w(a,b, c) \, ,
\end{array}
$$
by the definition of multiplication in $W$. Hence $\Psi$ induces a well-defined map on $\bw_{\KK }^3 H^1(X,\QQ)$:
\beq
\begin{array}{rrcl}
\tild\Psi : & \bw\limits_{\KK }^3 H^1(X,\QQ) & \larr & W\\
& a \wek b \wek c &\lmt & w(a,b,c) \, .
\end{array}
\eeq
Moreover, $\tild\Psi$ is surjective, since we can choose $a,b,c \in \{e_1,\dots, e_6\}$ to be distinct and the $\binom{6}{3}=20$ elements in the image give a $\KK$-basis of $W$. By comparing the dimensions of $W$ and $\bw_{\KK }^3 H^1(X,\QQ)$, we get that  $\tild{\Psi}$ gives the desired isomorphism.\\

\subsection{Splitting of $\bw_\KK^3 H^1(X,\QQ)$}
\noindent
Having established a $\KK$-linear isomorphism between the $\KK$-vector space $W$ and $\bw_\KK ^3 H^1(X,\QQ)$, we have to find a decomposition thereof, which induces a decomposition of $W$ into two Hodge substructures of type $(1,9,9,1)$. This is done in the same way as in section~\ref{sec:automorphism} on page~\pageref{sec:automorphism}. \\
We prove that there exists a natural $\KK$-antilinear automorphism $t$ of $\bw_\KK ^3 H^1(X,\QQ)$ such that $t \circ t = \rm{const} \cdot \rm{Id}$, whose eigenspaces give a decomposition of $\bw_\KK^3 H^1(X, \QQ)$. Each eigenspace will be a Hodge substructure of $W$.
We recall that $t$ is constructed as composition of  two linear maps. The first morphism makes use of the hermitian form $H$ \eqref{acca} restricted to $H^1(X, \QQ)  \simeq \KK^6 \subseteq \bCC^6$:
\beq H: H^1(X, \QQ) \times H^1(X, \QQ)  \larr  \KK \, ,
\eeq
which, on the standard basis $\{e_1,e_2, \dots, e_6\}$ of $\KK^6$, is given by $H(e_i,e_j) = \epsilon_i \delta_{ij}$. \\
$H$ can be extended to a form $\tild{H}$ on $\bw_{\KK }^3 H^1(X,\QQ)$, which is $\KK$-linear in the second variable and $\KK$-antilinear in the first variable, defined by
\beq
 \tild{H}( a\wek b\wek c, p\wek q\wek r)  = \det\left[
 \begin{array}{ccc}
 H(a,p) & H(a,q) & H(a,r) \\
 H(b,p) & H(b,q) & H(b,r) \\
 H(c,p) & H(c,q) & H(c,r)
 \end{array}
 \right] \, .
\eeq
Since $\tild{H}$ is $\KK$-linear in the second variable and $\KK$-antilinear in the first, it induces a $\KK$-antilinear map
\beq
\begin{array}{rrcl}
\tau: & \bw\limits_{\KK }^3 H^1(X,\QQ) & \larr &  \rm{Hom}_\KK \left(  \bw\limits_{\KK }^3 H^1(X,\QQ), \KK\right) \\
& \alpha & \lmt & \left[\beta \lmt \tild{H} (\alpha,\beta)\right] \, ,
\end{array}
\eeq
acting as follows:
\beq
e_i \wek e_j \wek e_k \stackrel{\tau}{\lmt} \epsilon_{ijk} (e_i \wek e_j \wek e_k )^* \, ,
\eeq
since $H(e_i,e_j) = \epsilon_i \delta_{ij}$ (here $(e_j \wek e_k \wek e_l )^*$ denotes the dual basis of $e_j \wek e_k \wek e_l$).
The second morphism is induced by the isomorphism
$$
\begin{array}{rrcl}
\gamma: &\bw\limits_{\KK }^3 H^1(X,\QQ)  \times \bw\limits_{\KK }^3 H^1(X,\QQ) &  \stackrel{\simeq}{\larr}  &  \bw\limits_{\KK }^6 H^1(X,\QQ)  \stackrel{\simeq}{\larr}  \KK  \\
&(\theta,\eta) &\lmt & \theta \wek \eta \, ,
\end{array}
$$
which sends $e_1\wek e_2 \wek e_3\wek e_4 \wek e_5\wek e_6$ to $1$. Hence we get an isomorphism
$$
\begin{array}{rrcl}
\rho: & \bw\limits_\KK ^3 H^1(X,\QQ) &\stackrel{\simeq}{\larr}& \rm{Hom}_{\KK }\left( \bw\limits_\KK^3 H^1(X,\QQ), \KK \right) \\
& \alpha  \lmt [ \beta & \lmt & \gamma(\alpha \wek \beta) ] \, ,
\end{array}
$$
which acts in the following way:
\beq
e_i \wek e_j \wek e_k  \stackrel{\rho}{\lmt} \delta_{ijk} (e_l \wek e_m \wek e_n )^*
\eeq
where $l,m,n \in \{1, \dots, 6\} \backslash \{i,j,k\}$ and $\delta_{ijk} =  \pm 1$.\\

\noindent
Using $\tau$ and $\rho$, we define the automorphism $t$:
\beq
t :=  \rho^{-1} \circ \tau: \bw\limits_\KK ^3 H^1(X,\QQ)  \larr  \bw\limits_\KK ^3 H^1(X,\QQ) \, .
\eeq
Obviously, since $\tau(w) = \rho(t(w))$, we have
\beq
\label{hgt1}
\tild{H}(v,w)= \gamma (t(w) \wek v) \, .
\eeq
Since $\rho$ is $\KK$-linear and $\tau$ $\KK$-antilinear, it follows that $t$ is $\KK$-antilinear. We can write the action of $t$ explicitly on the elements of the basis $\{e_i\wek e_j \wek e_k\}$, e.g.\
$$\begin{array}{ccccc}
e_1\wedge e_2\wedge e_3 & \lmt & (e_1\wedge e_2\wedge e_3)^*& \lmt &  -  e_4\wedge e_5\wedge e_6\\
 e_4\wedge e_5\wedge e_6 & \lmt & -(e_4\wedge e_5\wedge e_6)^* & \lmt & - e_1\wedge e_2\wedge e_3
\end{array}
$$
\\
\noindent
We found (see \ref{table1}) that $t^2= \rm{Id}$, and that $t$ has the eigenvalues $\pm1$, each with multiplicity $10$. Thus we have the decomposition:
\beq
W \simeq W_+ \oplus W_- \, , \quad \quad \dim W_+ =\dim W_- =10\, .
\eeq
\subsection{Hodge substructures of $W$}\noindent
Now we want to prove that $W_\pm$ are Hodge substructures of $W$. Taking $X$ as before with complex structure $J$ on $H^1(X, \bRR)$ as in section
\ref{B1}, we put
\beq
\label{rep1}
h:\bCC^*\longrightarrow GL(H^1(X, \bRR))\, , \quad h(a + b i) v := (aI +b J)v \, ,
\eeq
for all $a,b \in \bRR$, $v \in H^1(X, \bRR)$. Hence $h$ gives the scalar multiplication by complex numbers on $(H^1(X, \bRR), J)$. Then $H^{1,0}(X)$ and $H^{0,1}(X)$ are the eigenspaces of $h(z)$ with eigenvalues $z, \ov{z}$. \\
The representation \eqref{rep1}, associated to the Hodge structure of $H^1(X,\QQ)$, induces a representation on $W_\bRR \simeq \left( \bw_\KK^3 H^1(X,\QQ)\right) \otimes \bRR$,
\beq
h_3 := \bw^3 h : \bCC^*\larr GL( \Big( \bw_\KK^3 H^1(X,\QQ)\Big) \otimes \bRR)
\eeq
defined by
\beq
h_3(z) \left( u \wek v \wek w\right) :=  h(z) u \wek h(z) v \wek h(z) w \, ,
\eeq
which gives the Hodge structure on $W$.\\

\noindent
On $\bCC^6$ we have the action of $J$, which gives the decomposition:
\renewcommand{\arraystretch}{1.6}{$$
\begin{array}{rrccc}
(aI +bJ) & \circlearrowright  \bCC^6 =&  V_+ & \oplus & V_- \\
&& a+b i && a-bi \, .
\end{array}$$}
Hence
\renewcommand{\arraystretch}{1.6}{$$
\begin{array}{rrccc}
\bw\limits^6 \bCC^6 &= & \left( \bw\limits^3 V_+ \right) & \otimes & \left( \bw\limits^3 V_- \right) \\
&& (a+b i)^3 && (a-bi)^3 \, .
\end{array}$$}
Since $\left( \bw_\KK^6 \KK^6 \right) \otimes \bRR \simeq \bw^6 \bCC^6 \simeq \bCC$, we get that
$$ h_6(z) (\alpha \wek \beta) = h_3(z)\alpha \wek h_3(z) \beta = |z|^6 (\alpha \wek \beta) \quad \mbox{for all } \alpha,\beta \in W \, .
$$
\\
On the other hand, since $H(J v, J w) = H(v,w)$ (which implies $H(v,Jw)=H(Jv,J^2w)= -H(Jv,w)$)  and $h(a+b i)v=(aI +b J)v$, we also have that
$$
\begin{array}{rcl}
H(h(a+b i)v,h(a+b i)w) &=& H((aI +b J)v, (aI +b J)w) \\
&=& a^2 H(v,w)+ab\big(H(Jv,w)+H(v,Jw)\big)+b^2 H(v,w) \\
&=&  (a^2+b^2) H(v,w) \, .
\end{array}
$$
By the definition of $\tild{H}$, we get:
$$\tild{H}(h_3(z)\alpha, h_3(z) \beta) = (z \ov{z})^3 \tild{H}( \alpha,\beta ) =  |z|^6 \tild{H}( \alpha ,\beta )  \, .$$
Hence, using \eqref{hgt1}, we can conclude that
\beq
\begin{array}{rcl}
\gamma(t(h_3(z)\alpha) \wek h_3(z) \beta) &=& \tild{H}(h_3(z)\alpha, h_3(z) \beta) =  |z|^6 \tild{H}( \alpha,\beta )  = |z|^6  \gamma(t(\alpha) \wek \beta) \\
&=&  \gamma(h_3(z) t(\alpha) \wek h_3(z) \beta) \, ,
\end{array}
\eeq
for all $v,\beta \in W$, where in the last equality we used tha fact that $\gamma$ is $\KK$-bilinear and, once we take its $\bRR$-linear extension, it becomes $\bCC$-linear. Thus $t \circ h_3(z) = h_3(z) \circ t$, i.e.\ $t \in \rm{End}_{\rm{Hod}} (W)$. \\

\noindent
Since $t^2 = \rm{Id}$,
\beq W_\pm = \ker(t \pm \rm{Id})
\eeq
are Hodge substructures of $W$.\\
Moreover, there exists an isomorphism of Hodge structures given by
\beq
\Delta:= (2 \Theta + I) : W_+ \stackrel{\simeq}{\larr} W_- \quad \mbox{with } \Delta = 2 \Theta + I \, ,
\eeq
\\
\noindent
in fact notice that $\ov\Delta = 2\ov\Theta + I = 2 (-\Theta -I)+I= -(2 \Theta + I )= -\Delta$. If $w \in W_+$, we have that $t(w)=w$ and, since $t$ is $\KK$-antilinear, $t(\Delta w) = \ov\Delta t(w) = -\Delta w$. Hence $\Delta w \in W_-$.\\

\noindent
\subsection{Hodge structure of {\boldmath{$W_+$}} and {\boldmath{$SU(3,3)/S(U(3) \times U(3))$}}}
Recall that the group $G = SU(3,3)$ acts on the set of three-dimensional subspaces $V_+ \subset \bCC^6$. Now $V_+ \in \cS_G$ identifies the complex structure $J$ which in turn determines the Hodge structure of weight one associated to the representation $h(a+bi)$. Then we have the following correspondence:
\beq
g V_+ \longleftrightarrow g h(a + bi) g^{-1}.
\eeq
On $\bw^3 \KK^6 \otimes \bRR$ we have that
\beq
\bw^3 h(a+bi) =  h_3(a+bi)
\eeq
and
\beq
\bw^3 (g h(a+bi) g^{-1}) = g_3 h_3(a+bi) g_3^{-1}.
\eeq
Hence we have an action of $SU(3,3)$ on $W_+ \otimes \bRR = \bw^3 \KK^6 \otimes \bRR \simeq \bRR^{20}$ induced by the correspondence
\beq
g V_+ \longleftrightarrow g_3 h_3(a + bi) g_3^{-1} \, .
\eeq
Thus $SU(3,3)$ acts on the Hodge structures of weight $3$ on $\bw^3 \KK^6$. \\
Moreover, the stabilizer of $h_3(a+bi)$ is $S(U(3)\times U(3))$. Hence the moduli space of Hodge structures of weight $3$ on
$\bw^3 \KK^6$ is $SU(3,3)/S(U(3) \times U(3)$.

\newpage

\beq
\label{table1}
\tag{Table $1$}
\begin{array}{c|c|c|c}
& \tau & \rho &  \rho^{-1} \circ \tau  \\ \hline
e_1\wedge e_2\wedge e_3 & +   (e_1\wedge e_2\wedge e_3)^* & + (e_4\wedge e_5\wedge e_6)^* & -  e_4\wedge e_5\wedge e_6\\
\hline
 e_1\wedge e_2\wedge e_4 & -   (e_1\wedge e_2\wedge e_4)^* & - (e_3\wedge e_5\wedge e_6)^* &  -  e_3\wedge e_5\wedge e_6\\
 \hline
 e_1\wedge e_2\wedge e_5 & -   (e_1\wedge e_2\wedge e_5)^* & + (e_3\wedge e_4\wedge e_6)^* &  e_3\wedge e_4\wedge e_6 \\
 \hline
 e_1\wedge e_2\wedge e_6 & -   (e_1\wedge e_2\wedge e_6)^* & - (e_3\wedge e_4\wedge e_5)^*  & - e_3\wedge e_4\wedge e_5 \\
 \hline
 e_1\wedge e_3\wedge e_4 & -   (e_1\wedge e_3\wedge e_4)^* & + (e_2\wedge e_5\wedge e_6)^* &   e_2\wedge e_5\wedge e_6  \\
 \hline
 e_1\wedge e_3\wedge e_5 & -   (e_1\wedge e_3\wedge e_5)^* & - (e_2\wedge e_4\wedge e_6)^* &  -  e_2\wedge e_4\wedge e_6\\
 \hline
 e_1\wedge e_3\wedge e_6 & -   (e_1\wedge e_3\wedge e_6)^* & + (e_2\wedge e_4\wedge e_5)^* &  e_2\wedge e_4\wedge e_5 \\
 \hline
 e_1\wedge e_4\wedge e_5 & +   (e_1\wedge e_4\wedge e_5)^* & + (e_2\wedge e_3\wedge e_6)^*  & -  e_2\wedge e_3\wedge e_6 \\
 \hline
 e_1\wedge e_4\wedge e_6 & +   (e_1\wedge e_4\wedge e_6)^* & - (e_2\wedge e_3\wedge e_5)^*  &  e_2\wedge e_3\wedge e_5 \\
 \hline
 e_1\wedge e_5\wedge e_6 & +   (e_1\wedge e_5\wedge e_6)^* & + (e_2\wedge e_3\wedge e_4)^*  & - e_2\wedge e_3\wedge e_4 \\
 \hline
 e_2\wedge e_3\wedge e_4 & -   (e_2\wedge e_3\wedge e_4)^* & - (e_1\wedge e_5\wedge e_6)^*  &  -  e_1\wedge e_5\wedge e_6\\
 \hline
 e_2\wedge e_3\wedge e_5 & -   (e_2\wedge e_3\wedge e_5)^* & + (e_1\wedge e_4\wedge e_6)^*  &   e_1\wedge e_4\wedge e_6\\
 \hline
 e_2\wedge e_3\wedge e_6 & -   (e_2\wedge e_3\wedge e_6)^* & - (e_1\wedge e_4\wedge e_5)^* &   - e_1\wedge e_4 \wedge e_5\\
 \hline
 e_2\wedge e_4\wedge e_5 & +   (e_2\wedge e_4\wedge e_5)^* & - (e_1\wedge e_3\wedge e_6)^* & e_1\wedge e_3\wedge e_6 \\
 \hline
 e_2\wedge e_4\wedge e_6 & +   (e_2\wedge e_4\wedge e_6)^* & + (e_1\wedge e_3\wedge e_5)^* &  - e_1\wedge e_3\wedge e_5 \\
 \hline
 e_2\wedge e_5\wedge e_6 & +   (e_2\wedge e_5\wedge e_6)^* & - (e_1\wedge e_3\wedge e_4)^* &  e_1\wedge e_3\wedge e_4 \\
 \hline
 e_3\wedge e_4\wedge e_5 & +   (e_3\wedge e_4\wedge e_5)^* & + (e_1\wedge e_2\wedge e_6)^* &  - e_1\wedge e_2\wedge e_6 \\
 \hline
 e_3\wedge e_4\wedge e_6 & +   (e_3\wedge e_4\wedge e_6)^* & - (e_1\wedge e_2\wedge e_5)^* &  e_1\wedge e_2\wedge e_5\\
 \hline
 e_3\wedge e_5\wedge e_6 & +   (e_3\wedge e_5\wedge e_6)^* & + (e_1\wedge e_2\wedge e_4)^* &  - e_1\wedge e_2\wedge e_4  \\
 \hline
 e_4\wedge e_5\wedge e_6 & -   (e_4\wedge e_5\wedge e_6)^* & - (e_1\wedge e_2\wedge e_3)^* & - e_1\wedge e_2\wedge e_3
\end{array}
\eeq

\


\newpage


\begin{thebibliography}{99}

\bibitem[B]{B} F.~Bogomolov, {\it Hamiltonian K\"ahler manifolds}, Dokl. Akad. Nauk. SSSR{\bf 243} (1978), 1101--1104;
Soviet. Math. Dokl. {\bf 19} (1979), 1462--1465.

\bibitem[BB]{BB}
  V.~V.~Batyrev and L.~A.~Borisov,
  {\it Dual cones and mirror symmetry for generalized Calabi-Yau manifolds}, in ``Greene, B. (ed.): Yau, S.T. (ed.): Mirror symmetry II'', 71-86.

\bibitem[BG]{BG} R.L.~Bryant, P.A.~Griffths, {\it Some observations on the infinitesimal period relations for regular threefolds 
with trivial canonical bundle} in: Arithmetic and geometry, Vol. II, 77102, Progr. Math. 36, Birkhuser 
Boston, Boston, MA, 1983. 

\bibitem[BL]{BL} C.~Birkenhake, H.~Lange, {\it Complex Abelian varieties}, Springer-Verlag, Berlin (2004).

\bibitem[CDP]{CDP}
P.~Candelas, E.~Derrick and L.~Parkes,  {\it Generalized Calabi--Yau manifolds and the mirror of a rigid manifold},
Nucl.\ Phys.\  B {\bf 407} (1993) 115--154.

\bibitem[CGG]{CGG} J.~Carlson, M.~Green, Ph.~Griffths, {\it Variations of Hodge Structure Considered as an Exterior Differential 
System: Old and New Results}, SIGMA Symmetry Integrability Geom. Methods Appl. 5 (2009), Paper 
087. 
 
\bibitem[CHSW]{CHSW} P.~Candelas, G.~T.~Horowitz, A.~Strominger and E.~Witten,  {\it Vacuum Configurations For Superstrings},  Nucl.\ Phys.\  B {\bf 258} (1985) 46.

\bibitem[CvP]{Cremmer:1984hc} E.~Cremmer and A.~Van Proeyen, {\it Classification of K\"ahler manifolds in $N=2$ vector multiplet supergravity couplings},
Class. Quant. Grav. {\bf 2} (1985) 445.

\bibitem[CRTP]{CRTP} B.~Craps, F.~Roose, W.~Troost, A.~ Van Proeyen, {\it What is special K\"ahler geometry?}, Nucl.\ Phys.\ B {\bf 503} (1997) 565--613.
\bibitem[FFS]{FFS} S.~Ferrara, P.~Fr\'e and P.~Soriani, {\it On the moduli space of the $T^6/Z_3$ orbifold and its modular group},
Class. Quant. Grav. {\bf 9} (1992),1649--1662.

\bibitem[FLT]{FLT}  S.~Ferrara, D.~Lust and S.~Theisen, {\it Target Space Modular Invariance and Low-Energy Couplings in Orbifold  Compactifications}
Phys.\ Lett.\  B {\bf 233} (1989) 147.

\bibitem[Fr]{Fr} D.S.~Freed, {\it Special K\"ahler manifolds}, Comm. Math. Phys. {\bf 203} (1999), 31--52. 

\bibitem[vG]{vG} B.~van Geemen, \emph{An introduction to the Hodge conjecture for abelian varieties}, Algebraic cycles and Hodge theory (Torino, 1993), 233--252,
Lecture Notes in Math., 1594, Springer, Berlin, 1994.

\bibitem[GRV]{GRV} A.~Giveon, E.~Rabinovici and G.~Veneziano, {\it Duality in String Background Space} Nucl.\ Phys.\ B {\bf 322} (1989) 167--184.

\bibitem[Hel]{Hel} S.~Helgason, \emph{Differential geometry and symmetric spaces}, Academic Press (1962).

\bibitem[KLS]{KLS}
  S.~Kharel, M.~Lynker and R.~Schimmrigk,
  {\it String modular motives of mirrors of rigid Calabi-Yau varieties}, Fields Inst.\ Commun. {\bf 54} (2008) 47--63.

\bibitem[Lo]{Lo} G.~Lombardo, \emph{Abelian varieties of Weil type and Kuga-Satake varieties}, Tohoku Math. J. (2) {\bf 53} (2001), no. 3, 453--466.

\bibitem[Sch]{Sch}
R.~Schimmrigk, {\it Mirror symmetry and string vacua from a special class of Fano varieties}, Int.\ J.\ Mod.\ Phys.\ A {\bf 11} (1996) 3049.

\bibitem[Se]{Se}  S.~Sethi,
  {\it Supermanifolds, rigid manifolds and mirror symmetry},  Nucl.\ Phys.\ B {\bf 430} (1994) 31.

\bibitem[Sh]{Sh} D.~Shevitz, {\it The Global Structure of a $c=9$ Type $(2,2)$ Moduli Space} Nucl.\ Phys.\ B {\bf 338} (1990) 283--293.

\bibitem[Stro]{Stro} A.~Strominger, {\it Special geometry} Comm. Math. Phys. {\bf 133} (1990), 163--180.


\bibitem[Ti]{Ti} G.~Tian, {\it Smoothness of the universal deformation space of compact Calabi-Yau manifolds and its Petersson-Weil metric},
Mathematical Aspects of String Theory, San Diego (1986), Adv. Ser. Math. Phys. 1, World Scientific Publishing, Singapore, (1987), 629--646.

\bibitem[To]{To} A.~N.~Todorov, {\it The Weil-Petersson geometry of the moduli space of $SU(n\geq 3)$ (Calabi-Yau) Manifolds I},
Commun. Math. Phys. {\bf 126} (1989), 325--346.

\end{thebibliography}
\end{document}